\documentclass[prd,amsmath,amssymb,cite]{revtex4} 

\usepackage[utf8]{inputenc}
\usepackage{floatpag}
\usepackage{amsthm}
\usepackage{enumerate}
\usepackage{amssymb}
\usepackage{amsmath}
\usepackage{braket}
\usepackage{hyperref}
\usepackage{graphicx}
\usepackage{epstopdf}
\usepackage{slashed}
\usepackage{color} 

\def\beq{\begin{equation}}
\def\eeq{\end{equation}}
\def\beq{\begin{eqnarray}}
\def\eeq{\end{eqnarray}}
\def\tr{{\rm tr}}

\allowdisplaybreaks

\begin{document}

\title{Non-eikonal corrections to multi-particle production in the Color Glass Condensate}

\author{Pedro Agostini$^{a}$, Tolga Altinoluk$^{b}$ and N\'estor Armesto$^{a}$}

\affiliation{
$^a$ Instituto Galego de F\'{\i}sica de Altas Enerx\'{\i}as IGFAE, Universidade de Santiago de Compostela, 15782 Santiago de Compostela, Galicia-Spain \\
$^b$ National Centre for Nuclear Research, 00-681 Warsaw, Poland 
}

\begin{abstract}
 We consider the non-eikonal corrections to particle production in the Color Glass Condensate stemming from the relaxation of the shockwave approximation for the target that acquires a finite longitudinal dimension. We derive a modified expression of the Lipatov vertex which takes into account this finite target width. This expression is employed to compute single, double and triple gluon production in the Glasma graph limit valid for the scattering of two dilute objects, at all orders in the expansion in the number of colors.
 We justify and generalize previous results, and discuss the possible implications on two particle correlations of these non-eikonal corrections  that induce differences between the away- and near-side peaks.
\end{abstract}

\maketitle

\section{Introduction}
\label{sec:intro}

Particle production at high energies in the soft and semihard regimes is usually computed resourcing to high energy approximations~\cite{Kovchegov:2012mbw}, namely the eikonal approximation. This is the case in the Color Glass Condensate (CGC) \cite{Iancu:2002xk,McLerran:2008uj,Gelis:2010nm}. In this framework, the process of propagation of an energetic parton from the projectile through the target, considered as a background field, is computed in the light cone gauge neglecting its transverse components and considering it as infinitely time dilated and Lorentz contracted (thus treated as a shockwave), see for example the discussion in~\cite{Altinoluk:2015gia}. Also terms subleading in energy (among them, spin flip ones) are neglected. On the other hand, in the calculation of elastic and radiative energy loss of energetic partons traversing a medium composed of coloured scattering centers -- jet quenching -- the shockwave approximation is relaxed and the target is considered to have a finite length, see e.g. the reviews \cite{Kovner:2003zj,CasalderreySolana:2007pr}~\footnote{The relation between jet quenching and CGC calculations, using the formalism in \cite{Blaizot:2004wu,Gelis:2005pt}, was established in~\cite{MehtarTani:2006xq} where the validity of the eikonal approximation for this type of computations was also addressed.}. In this context, a systematic expansion of the gluon propagator in non-eikonal terms was done in~\cite{Altinoluk:2014oxa,Altinoluk:2015gia} and applied to particle production in the CGC in~\cite{Altinoluk:2015xuy}. Non-eikonal corrections at high energies have also been treated recently in the context of Transverse Momentum Distributions and spin physics~~\cite{Balitsky:2015qba,Balitsky:2016dgz,Kovchegov:2015pbl,Kovchegov:2016zex,Kovchegov:2017jxc,Kovchegov:2017lsr,Chirilli:2018kkw}, and soft gluon exponentiation~\cite{Laenen:2008ux,Laenen:2008gt,Laenen:2010uz}.

In the CGC, particle production and correlations have been computed within several approximation schemes, providing an alternative explanation to final state interactions for the ridge phenomenon observed in small systems, proton-proton and proton-nucleus, at the Large Hadron Collider (LHC) at CERN~\cite{Khachatryan:2010gv,Khachatryan:2015lva,Aad:2015gqa,CMS:2012qk,Abelev:2012ola,Aad:2012gla,Aaij:2015qcq,Khachatryan:2016ibd,Khachatryan:2016txc,Aaboud:2016yar,Aaboud:2017acw,Aaboud:2017blb,Chatrchyan:2013nka,Abelev:2014mda} and the Relativistic Heavy Ion Collider (RHIC) at BNL~\cite{Alver:2009id,Abelev:2009af,Adare:2014keg,Adamczyk:2015xjc,Adare:2015ctn}. The ``Glasma graph" approximation~\cite{Dumitru:2008wn,Dumitru:2010iy}, suitable for collisions between two dilute objects like proton-proton and containing both Bose enhancement and Hanbury-Brown-Twiss effects~\cite{Kovchegov:2012nd,Kovchegov:2013ewa,Altinoluk:2015uaa,Altinoluk:2015eka}, has been used to describe experimental data~\cite{Dusling:2012iga,Dusling:2012cg,Dusling:2012wy,Dusling:2013qoz}, and to compute three and four gluon correlations~\cite{Ozonder:2014sra,Ozonder:2017wmh}. Quark correlations have also been calculated in this framework~\cite{Altinoluk:2016vax,Martinez:2018ygo}. It was later extended to dilute-dense (proton-nucleus) collisions both numerically~\cite{Lappi:2015vta} and analytically~\cite{Altinoluk:2018hcu,Altinoluk:2018ogz,Martinez:2018tuf}, and used to calculate three gluon correlations~\cite{Altinoluk:2018ogz}. A description of data has been obtained~\cite{Dusling:2017dqg,Dusling:2017aot}. Density gradients~\cite{Levin:2011fb} have also been considered to explain the observed azimuthal structure. 

Beyond the analytical extension to dense-dense collisions, the remaining key theoretical problem for the description of azimuthal structure in small systems in the CGC lies in odd harmonics that are absent in usual calculations. For this, density corrections in the projectile~\cite{McLerran:2016snu,Kovner:2016jfp,Kovchegov:2018jun}, quark correlations~\cite{Dumitru:2014vka,Kovner:2017gab,Dusling:2017dqg} and a more involved description of the target~\cite{Kovner:2012jm,Dumitru:2014yza} than the one provided by the commonly used McLerran-Venugopalan (MV) model~\cite{McLerran:1993ni,McLerran:1994vd}, have been proposed. Using the former, a  description of data is possible~\cite{Mace:2018vwq,Mace:2018yvl,Mace:2019rtt}.

In this manuscript we deal with non-eikonal corrections to particle production in the CGC that stem from relaxing the shockwave approximation for the target, which becomes of finite length. These are the corrections included in jet quenching calculations and systematically expanded up to next-to-next-to-leading order in~\cite{Altinoluk:2014oxa,Altinoluk:2015gia}. In Section~\ref{sec:neLip} we derive an expression for the Lipatov vertex -- one central building block for particle production calculations in the CGC -- that takes into account the finite longitudinal extent of the target field.
While by itself this result is not new and similar calculations and expressions can be found in the literature, see e.g. Refs.~\cite{Wiedemann:2000za,Gyulassy:2000er} or more recently  in Ref.~\cite{MehtarTani:2011gf}, its identification for use to include non-eikonal corrections in CGC calculations is done here for the first time. Then, in Section~\ref{sec:multi} we apply our corrections to gluon production in the dilute-dilute (Glasma graph) limit, following the notations in~\cite{Altinoluk:2018ogz}. First, in Subsecion~\ref{sec:single} we consider single gluon production, matching the results in~\cite{Altinoluk:2015xuy} and justifying the educated guess done there on the basis of the expansion up to next-to-next-to-leading order. Then we consider double gluon production in Subsection~\ref{sec:double}, where we generalize the results in~\cite{Altinoluk:2015xuy}. Third, in Subsection~\ref{sec:triple} we compute three gluon production. Finally, in Section~\ref{sec:conclu} we discuss our results. 
We focus on providing analytical expressions and show a few numerical results; a more complete study of the impact of non-eikonal corrections on particle correlations is left for a forthcoming study~\cite{Azimuthal_Harm_NonEik}.

\section{Derivation of the non-eikonal Lipatov vertex}
\label{sec:neLip}

As usually done in the CGC, we describe a high energy p-A collision by a right moving dilute projectile which interacts with a left moving dense target described by a random and intense ($\mathcal{O}(1/g)$) classical gluon field $A^\mu(x)$. The simplest setup to derive the non-eikonal Lipatov vertex is considering the emission of a gluon from a projectile massless quark in the process of a single scattering with the target (an analogous calculation leading to the same conclusions on the non-eikonal corrections holds for a projectile gluon). In light cone coordinates $a^{\pm}=(a_0\pm a_3)/\sqrt{2}$ and in the light cone gauge ($n\cdot A=A^+=0$, $n=(0,1,0_\perp)$ in $(+,-,\perp)$ coordinates), this field can be written as
\begin{align} \label{eq1}
A^\mu(x)\approx \delta^{\mu -} \delta(x^+) A^-(x_\perp),
\end{align}
since the transverse component of the gluon field is not altered by the large Lorentz $\gamma$ factor, the $x^-$ dependence disappears due to the time dilatation and the target is shrinked to $x^+=0$ forming a shock-wave. However, in some applications these suppressed terms may be sizeable. For this reason, in this note we will relax the infinite boost approximation, in order to calculate the corresponding non-eikonal corrections to the usual Lipatov vertex computed at $\mathcal{O}(g^2)$.

\begin{figure}[h!]
	\centering
	\includegraphics[scale=0.5]{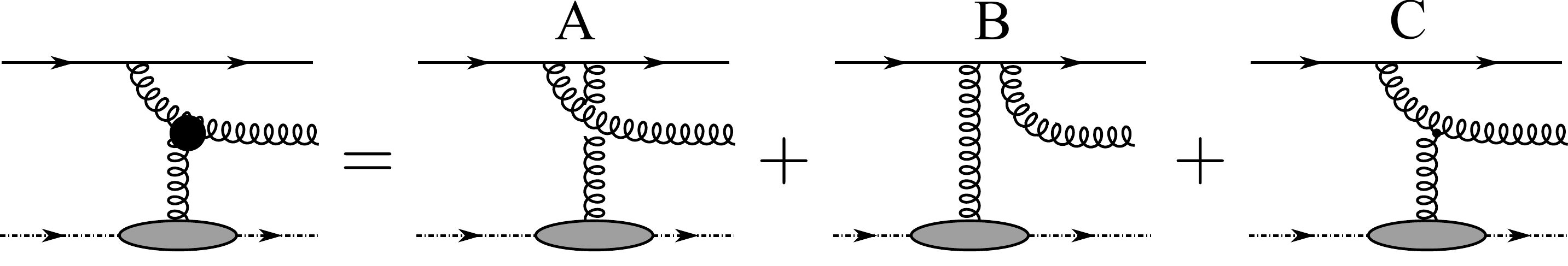}
	\caption{Diagrams that contribute to the computation of the Lipatov vertex. The black dot represents the Lipatov vertex which is the sum of all real diagrams for gluon production shown on the right hand side of the equation.}
	\label{fig1}
\end{figure}

To proceed, we analyze gluon production in p-A collisions in the quark initiated channel and compute the Lipatov vertex, which is an effective vertex that takes into account all the real contributions to gluon production. For that one needs to sum the amplitudes where the gluon is emitted before, during and after the interaction with the field as shown in Fig.~\ref{fig1}.

Our setup is such that the right moving quark with momentum $p+k-q$ is generated by some function $J(p+k-q)=J(p^++k^+-q^+)$  at $x_0^+=-\infty$ and $(x_0^-,x_{0\perp})=0$, and then interacts with the classical gluon field $A^\mu(x)$ generated by one scattering source located at $x_1$, picking up a momentum $q$. However, since we are interested in non-eikonal corrections, we consider $A^\mu(x)$ with an $x^+$ dependence which has a finite support instead of treating it as a shockwave at $x^+=0$, but we still assume that there is no dependence on $x^-$. That is, the new form of Eq.~(\ref{eq1}) is  
\begin{align}
A^\mu(x)\approx \delta^{\mu -} A^\mu(x^+,x_\perp),
\end{align}
or, in momentum space,
\begin{align}\label{eq3}
A^\mu(q)\approx \delta^{\mu -} \,2 \pi \delta(q^+) A^-(q^-,q_\perp).
\end{align}
Furthermore, we assume that the outgoing quark has a large momentum $p^+$ compared to all other momenta in the process. The general strategy in this case is to keep the leading terms in $+$-momenta in the numerator algebra, while taking the full phase corrections coming from the integration of the denominators, see below, as done in the Furry approximation and its non-abelian generalization \cite{Wiedemann:2000ez}.

\begin{figure}[h!]
	\centering
	\includegraphics[scale=0.7]{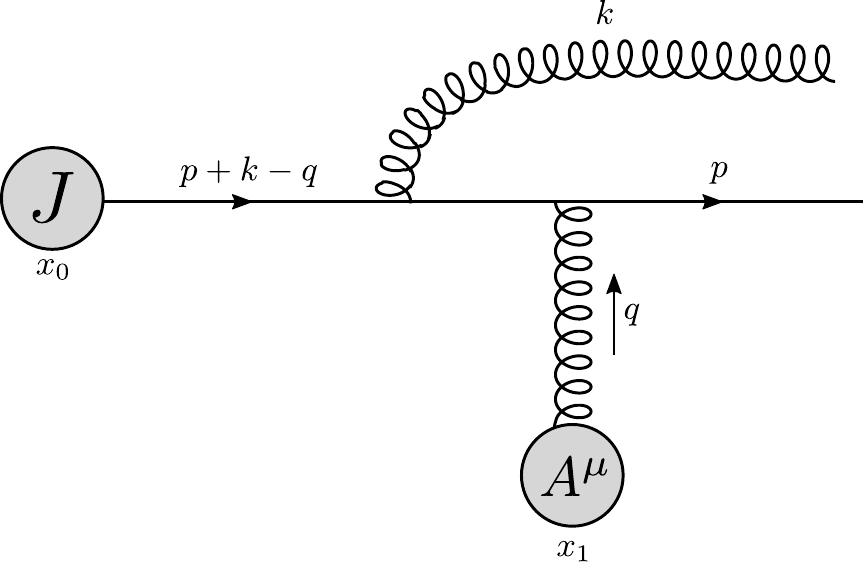}
	\caption{Diagram A where the gluon is emitted before the interaction of the quark  with the target field. }
	\label{fig2}
\end{figure}

We start by computing  diagram A where the gluon is emitted with momentum $k$ before the quark interaction with the target field as shown in Fig.~\ref{fig2}. Using the Feynman rules, we find that the amplitude for  fixed gluon and final quark momenta is
\begin{align}
i \mathcal{M}_A=&\bar{u}(p)(-ig\gamma^\mu t^a)\int \frac{d^4q}{(2 \pi)^4}A_\mu^a(q)e^{i q x_1} \frac{i (\slashed{p}-\slashed{q}) }{(p-q)^2+i \epsilon} (-i g \gamma^\nu t^b) \epsilon_\nu^{b *}(k) \nonumber \\
&\times \frac{i(\slashed{p}+\slashed{k}-\slashed{q})}{(p+k-q)^2+i\epsilon}e^{i(p+k-q)x_0} J(p+k-q),
\end{align}
with $t^a$ the $SU(N_c)$ generators in the fundamental representation.

Since $p^+$ is the largest momentum in our problem, we approximate $\slashed{p}-\slashed{q}\approx \slashed{p}$ and $\slashed{p}+\slashed{k}-\slashed{q} \approx \slashed{p}$ and write
\begin{align}
i \mathcal{M}_A\approx \bar{u}(p)e^{i(p+k)x_0} g^2 t^a t^b \int \frac{d^4q}{(2 \pi)^4} \frac{\slashed{A}^a(q) \slashed{p} \slashed{\epsilon}^{b *}(k) \slashed{p}}{[(p-q)^2+i \epsilon][(p+k-q)^2+i\epsilon]} e^{iq(x_1-x_0)}J(p^++k^+-q^+).
\end{align}

Using again the eikonal approximation ($p^+$ much larger than all other momenta), we can approximate $(p-q)^2 \approx -2 p^+ q^-$ and $(p+k-q)^2 \approx 2 p^+(k^- - q^-)$. Employing $\slashed{a} \slashed{b}=2 a\cdot b - \slashed{b} \slashed{a}$ and the massless Dirac equation $\bar{u}(p) \slashed{p}=0$, we get $\bar{u}(p) \slashed{A}^a(q) \slashed{p} \slashed{\epsilon}^{b *}(k) \slashed{p}=\bar{u}(p) 4 (p\cdot A^a(q)) (p \cdot \epsilon^{b *}(k))$. Therefore, the amplitude for diagram A can be written as 
\begin{align}
i \mathcal{M}_A\approx&-\bar{u}(p)e^{i(p+k)x_0} g^2 t^a t^b \int \frac{d^4q}{(2 \pi)^4} \frac{(p\cdot A^a(q)) (p \cdot \epsilon^{b *}(k))}{[p^+ q^- -i \epsilon][p^+(k^- - q^-)+i\epsilon]} e^{iq (x_1-x_0)} J(p^++k^+-q^+) \nonumber \\
=&-\bar{u}(p)e^{i(p+k)^- x_0^+} g^2 t^a t^b \int \frac{d^2 q_\perp}{(2 \pi)^2}e^{-iq_\perp x_{1 \perp}} (p \cdot \epsilon^{b *}(k)) \int \frac{dq^+}{2\pi}e^{iq^+ x_1^-} J(p^++k^+-q^+) (2 \pi) \delta(q^+)\nonumber \\
&\times \int \frac{dq^-}{2 \pi}\frac{e^{iq^- (x_1^+-x_0^+)} p^+ A^{-a}(q^-,q_\perp)}{(p^+)^2[q^- -i \epsilon][k^- - q^- +i\epsilon]},
\end{align}
where in the last line we used Eq.~(\ref{eq3}) and we have set $x_{0\perp}=x_0^-=0$. Performing the $q^+$ and $q^-$ integrals we obtain 
\begin{align}
i \mathcal{M}_A\approx&-\bar{u}(p) e^{ip x_0 }J(p^++k^+) g^2 t^a t^b \int \frac{d^2 q_\perp}{(2 \pi)^2}e^{-iq_\perp x_{1 \perp}} p \cdot \epsilon^{b *}(k) \nonumber \\
&\times \frac{i\left[e^{ik^- x_0^+}A^{-a}(0,q_\perp)-e^{i k^- x_1^+}A^{-a}(k^-,q_\perp)\right]}{p^+ k^-} \ \Theta(x_1^+-x_0^+).
\end{align}

Since the outgoing gluon is on-shell, $k^-=k_\perp^2/2k^+$ and, furthermore, in the light cone gauge we have $\epsilon^{*-}(k)=k^i \epsilon^i/k^+$. Therefore, making use of $p^\mu \epsilon_\mu^* \approx p^+ \epsilon^{*-}$, we obtain 
\begin{align}
i \mathcal{M}_A&\approx 2i\bar{u}(p) e^{ipx_0} J(p^++k^+) g^2 t^a t^b \Theta(x_1^+-x_0^+)  \frac{k^i \epsilon^{bi}}{k_\perp^2} \nonumber \\ 
&\times \int \frac{d^2 q_\perp}{(2 \pi)^2}e^{-iq_\perp x_{1 \perp}} \left(e^{i k^- x_1^+}A^{-a}(k^-,q_\perp)-e^{ik^- x_0^+}A^{-a}(0,q_\perp)\right).
\end{align}

Now, sending $x_0^+\rightarrow - \infty$ we can finally write 
\begin{align}
i \mathcal{M}_A\approx 2i\bar{u}(p) e^{ipx_0} J(p^++k^+) g^2 t^a t^b e^{i k^- x_1^+} \frac{k^i \epsilon^{bi}}{k_\perp^2}  \int \frac{d^2 q_\perp}{(2 \pi)^2}e^{-iq_\perp x_{1 \perp}} A^{-a}(k^-,q_\perp) .
\end{align}

\begin{figure}[h!]
	\centering
	\includegraphics[scale=0.7]{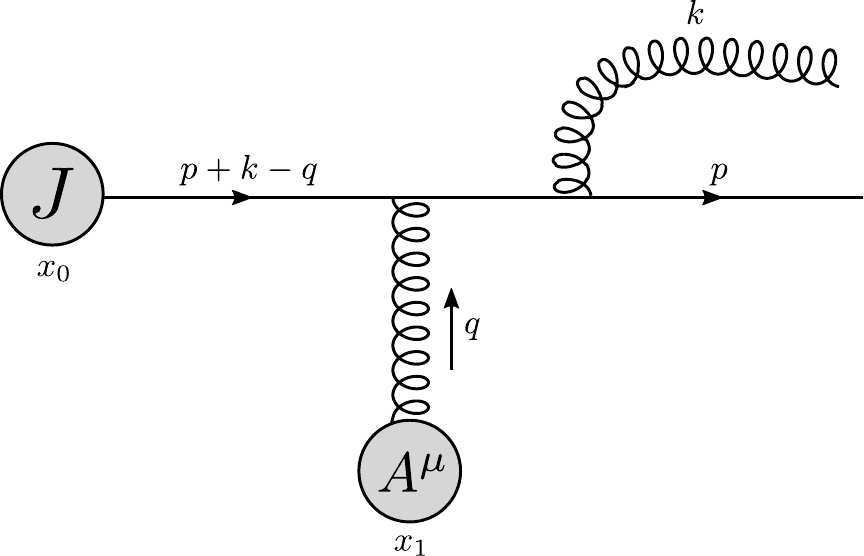}
	\caption{Diagram B where the gluon is emitted after the  interaction of the quark with the target field. }
	\label{fig3}
\end{figure}

Now we proceed to calculate diagram B where the gluon is emitted with momentum $k$ after the  interaction of the quark with the target field, as shown in Fig.~\ref{fig3}. Following the previous procedure we find 
\begin{align}
i \mathcal{M}_B\approx -2i\bar{u}(p) e^{i p x_0} J(p^++k^+) g^2 t^b t^a e^{i k^- x_1^+} \frac{k^i \epsilon^{bi}}{k_\perp^2}  \int \frac{d^2 q_\perp}{(2 \pi)^2}e^{-iq_\perp x_{1 \perp}} A^{-a}(k^-,q_\perp) .
\end{align}

\begin{figure}[h!]
	\centering
	\includegraphics[scale=0.7]{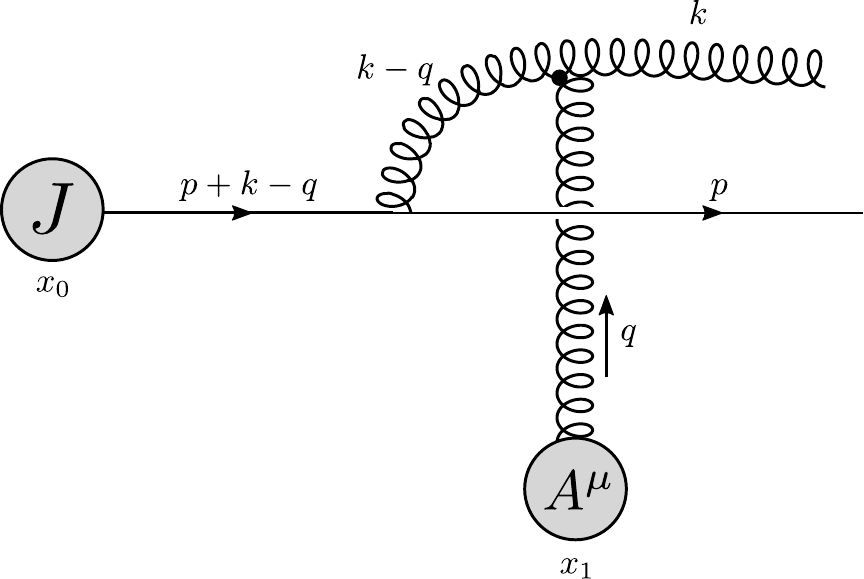}
	\caption{Diagram C where the emitted gluon interacts with the target field. }
	\label{fig4}
\end{figure}

Diagram C, shown in Fig.~\ref{fig4}, where the emitted gluon interacts with the target field, requires dealing with the three-gluon vertex. Applying the Feynman rules we have 
\begin{align}
i \mathcal{M}_C=&\bar{u}(p)(-i g \gamma^{\mu} t^{a}) \int \frac{d^4 q}{(2\pi)^4}\frac{i(\slashed{p}+\slashed{k}-\slashed{q})}{(p+k-q)^2+i\epsilon} e^{i(p+k-q)x_{0}} J(p+k-q) \nonumber 
\\ &\times \frac{-i d_{\mu \alpha}(k-q)}{(k-q)^2+i\epsilon}V^{\alpha \nu \beta}_{abc}A_{\beta}^{c}(q)\epsilon_{\nu}^{b *}(k) e^{i q x_1},
\end{align}
where $V^{\alpha \nu \beta}_{abc}=g f^{abc} \left[ g^{\alpha \nu} (q-2k)^\beta+g^{\nu \beta} (k+q)^\alpha+g^{\beta \alpha} (k-2q)^\nu \right]$ is the three-gluon vertex and $d_{\mu \alpha}(k)=g_{\mu \alpha}-\frac{k_\mu n_\alpha+k_\alpha n_\mu}{k \cdot n}$ the gluon propagator in the light cone gauge.

Considering Eq.~(\ref{eq3}), we  only need the $V^{\alpha \nu +}_{abc}$ component of the vertex. Furthermore, using the Dirac equation and the gamma matrices anti-commutation relation we have that $\bar{u}(p)\gamma^\mu \slashed{p}=\bar{u}(p) 2p^\mu$. Thus,
\begin{align}
i \mathcal{M}_C=&-2i\bar{u}(p) e^{i(p+k)x_{0}} g^2 t^{a} \int \frac{d^4 q}{(2\pi)^4}\frac{p^\mu d_{\mu \alpha} (k-q)V^{\alpha \nu +}_{abc}\epsilon_\nu^{b*}(k)}{[(p+k-q)^2+i\epsilon][(k-q)^2+i\epsilon]} A^{-c}(q) e^{iq(x_1-x_0)} J(p^++k^+-q^+).
\end{align}

After some algebra we find, in the eikonal approximation,
\begin{align}
p^\mu d_{\mu \alpha} (k-q)V^{\alpha \nu +}_{abc}\epsilon_\nu^{b*}(k)\approx -2gf^{abc}p^+(k-q)^i\cdot \epsilon^{bi},
\end{align}
and 
\begin{align}
(k-q)^2= -2k^+\left( q^- - \frac{k_\perp^2-(k-q)_\perp^2}{2k^+}  \right) +2q^+ q^- -2k^- q^+ \approx -2k^+\left( q^- - \frac{k_\perp^2-(k-q)_\perp^2}{2k^+}  \right).
\end{align}

Thus, defining $\tilde{k}=\frac{k_\perp^2-(k-q)_\perp^2}{2k^+}$, we get
\begin{align}
i \mathcal{M}_C\approx -i\bar{u}(p) g^2 t^af^{abc} e^{i(p+k)x_0}\int \frac{d^4 q}{(2\pi)^4}(k-q)^i \epsilon^{bi} \ \frac{e^{iq(x_1-x_0)}}{k^+[k^--q^-+i\epsilon][q^--\tilde{k}-i\epsilon]} A^{-c}(q) J(p^++k^+-q^+). 
\end{align}

Using Eq.~(\ref{eq3}) and performing the $q^+$ and $q^-$ integrals we obtain
\begin{align}
i \mathcal{M}_C\approx &2 \bar{u}(p)J(p^++k^+) e^{ipx_0} g^2 t^a f^{abc}\int \frac{d^2 q_\perp}{(2\pi)^2}\frac{(k-q)^i}{(k-q)_\perp^2} \epsilon^{bi}  e^{-iq_\perp x_{1\perp}} \nonumber \\
&\times e^{i \tilde{k} x_1^+} \left(  e^{i(k^--\tilde{k})x_0^+} A^{-c}(\tilde{k},q_\perp)- e^{i(k^--\tilde{k})x_1^+} A^{-c}(k^-,q_\perp) \right) \Theta (x_1^+-x_0^+).
\end{align}

 Finally, making use of $i t^a f^{abc}=[t^b,t^c]$ and sending $x_0^+ \rightarrow - \infty$, we obtain
 \begin{align}
 i \mathcal{M}_C\approx - 2 i \bar{u}(p)J(p^++k^+) e^{ipx_0} g^2 [t^a,t^b] \int \frac{d^2 q_\perp}{(2\pi)^2}\frac{(k-q)^i}{(k-q)_\perp^2} \epsilon^{bi}  e^{ik^- x_1^+}  A^{-a}(k^-,q_\perp)e^{-iq_\perp x_{1\perp}}.
 \end{align}
 
 Summing up the three diagrams we get
 \begin{align}
& i(\mathcal{M}_A+\mathcal{M}_B+\mathcal{M}_C) \nonumber \\
&\approx- 2 i \bar{u}(p)J(p^++k^+) e^{ipx_0} g^2 [t^a,t^b] \int \frac{d^2 q_\perp}{(2\pi)^2} L^i(k_\perp,q_\perp) \epsilon^{bi}  e^{ik^- x_1^+}  A^{-a}(k^-,q_\perp)e^{-iq_\perp x_{1\perp}},
 \end{align}
 where
 \begin{equation}
 L^i(k_\perp,q_\perp)=\frac{(k-q)^i}{(k-q)_\perp^2}-\frac{k^i}{k_\perp^2}
 \label{EikL}
 \end{equation}
 is the eikonal Lipatov vertex. We see that in our calculation, as announced, the non-eikonal corrections result in the sum of the amplitudes simply picking up a phase (important for  $k^-x_1^+ \sim 1$ with $k^-\propto k_\perp^2 e^{-\eta}$ and negligible for $k_\perp^2 x_1^+/k^+\ll 1$ where we recover the eikonal result) that can be absorbed in a redefinition of the Lipatov vertex. Therefore, we  define a non-eikonal Lipatov vertex
  \begin{align}
 L_{\text{NE}}^i(\underline{k},q_\perp;x_1^+)=\left[ \frac{(k-q)^i}{(k-q)_\perp^2}-\frac{k^i}{k_\perp^2} \right] e^{i\frac{k_\perp^2}{2 k^+} x_1^+},
 \label{NEikL}
 \end{align}
 with $\underline{k}\equiv (k^-,k_\perp)$.
 
 As stated in the Introduction, this result is not new by itself and similar calculations and expressions can be found in the literature, e.g. in Refs.~\cite{Wiedemann:2000za,Gyulassy:2000er} or later in Ref.~\cite{MehtarTani:2011gf}. But the identification of this building block for its use to include non-eikonal corrections in CGC calculations is done here for the first time.
Note that using the non-eikonal expression of the gluon propagator from~\cite{Altinoluk:2014oxa,Altinoluk:2015gia}, the two first terms of the expansion of the exponential were obtained in~\cite{Altinoluk:2015xuy} and the exponential form guessed.

\section{Multi-particle production}
\label{sec:multi}

In the previous section, we have presented the derivation of the non-eikonal Lipatov vertex. Now, we would like to use this expression in order to calculate multi-gluon production cross section at mid rapidity within the Glasma graph approach in order to study the effects of finite target width corrections to those observables.

The double and triple inclusive gluon production cross sections in p-A collisions have been recently studied in~\cite{Ozonder:2014sra,Ozonder:2017wmh} in the Glasma graph approximation, and in~\cite{Altinoluk:2018ogz} going beyond it, i.e. taking into account multiple scattering effects of the dense target. For each observable, the contributions to Bose enhancement of the projectile gluons and HBT contributions of the final state gluons are identified. However, the studies in~\cite{Ozonder:2014sra,Ozonder:2017wmh,Altinoluk:2018ogz} are performed within the eikonal approximation without taking into account the corrections due to the finite longitudinal width of the target.

In the rest of this section, we take this extra step. Namely, we first expand the single, double and triple inclusive gluon production cross section in powers of the background field of the target which actually corresponds to the original Glasma graph approach. Then, we introduce the non-eikonal Lipatov vertex \eqref{NEikL} in the expanded cross sections and get the explicit expressions of the Bose enhancement and HBT contributions beyond the strict eikonal limit for the double and triple inclusive gluon production. Hereafter, in order to alleviate the notation we will drop the $\perp$ for denoting transverse coordinates and momenta.

\subsection{Single inclusive gluon production beyond the eikonal approximation}
\label{sec:single}

Within the CGC framework, the production cross section of a gluon with transverse momenta $k$ and rapidity $\eta$ can be written  
\begin{equation}
\label{xsection}
\frac{d\sigma}{d^2 k d\eta}= 4 \pi \alpha_s \int_{z \bar{z}} e^{i k (z-\bar{z})} \int_{x y} A^i(x-z)A^i(\bar{z}-y) \Big\langle \rho^a(x) \rho^b(y)\Big\rangle_P 
\Big\langle\big[U_z-U_x\big]^{ac} \big[ U^\dagger_{\bar{z}}-U^\dagger_y \big]^{cb}\Big\rangle_T,
\end{equation}
where $\rho^a(x)\equiv \rho_x^a$ is the colour charge density of the projectile, $\langle \cdots \rangle_{P(T)}$ denote the average over the projectile (target) colour configurations and $A^i$ is the standard Weiz\"acker-Williams field that is defined as 
\begin{equation}
A^i(x-y)=-\frac{1}{2\pi}\frac{(x-y)^i}{(x-y)^2}=\int \frac{d^2p}{(2\pi)^2}\, e^{-ip\cdot(x-y)}\frac{p^i}{p^2}\ .
\end{equation}
Moreover, we have introduced a short hand notation for the transverse coordinate integrals $\int_z=\int d^2z$. Here, $U_x^{ab}$ is the adjoint Wilson line in the colour field of the target representing the scattering matrix of a gluon at transverse position $x$, whose explicit expression reads
\beq
U^{ab}_x={\cal P}\, e^{ig\int dx^+ T^c_{ab}\,A^-_c(x^+,x)},
\eeq
with $T^c_{ab}$ being the $SU(N_c)$ generator in the adjoint representation and  $A^-_c(x^+,x)$  the colour field of the target. The Wilson line operator accounts for the multiple scattering effects of the gluon in its interaction with the target. However, as mentioned previously, the Glasma graph approach for double (or multiple) gluon production corresponds to the dilute limit of the target. Therefore, we expand the Wilson lines to first order in the colour field of the target:
\begin{align}
\label{expanded_U}
U_{ab}(x) \approx 1+i g T^c_{ab} \int dx^+  A^-_c(x^+,x)= 1+i g T^c_{ab} \int dx^+ \int \frac{d^2q}{(2\pi)^2}\,  e^{i q x} \,  A^-_c(x^+,q)\, .
\end{align}
Using Eq.~\eqref{expanded_U} we can write the single inclusive gluon production cross section in the dilute limit as 
\beq
\label{Dilute_Single_X_sec}
&&
\frac{d\sigma}{d^2 k d\eta}\bigg|_{\rm dilute}= 4 \pi \alpha_s \int_{z \bar{z}} e^{i k (z-\bar{z})} \int_{x y} A^i(x-z)A^i(\bar{z}-y) \Braket{\rho^a(x) \rho^b(y)}_P \nonumber \\
&&
\times g^2 \int dx_1^+ \, dx_2^+ \int \frac{d^2q_1}{(2\pi)^2} \, \frac{d^2q_2}{(2\pi)^2}  \Big\langle A^-_{c}(x_1^+,q_1) A^-_{\bar{c}}(x_2^+,q_2)\Big\rangle_T (T^c T^{\bar{c}})_{ab} \left[e^{-iq_1 \cdot z}-e^{-iq_1\cdot  x}\right] \left[e^{iq_2 \cdot \bar{z}}-e^{iq_2\cdot y}\right].
\eeq
We can now perform the colour averaging over the projectile colour charge densities. For the correlator of two projectile colour charge densities, we use the generalized MV model and write it in the following general form:
%
\begin{align}
\label{proj_corr}
\Braket{\rho^a(x) \rho^b(y)}_P=\delta^{ab} \, \mu^2(x,y).
\end{align}
Inserting Eq.~\eqref{proj_corr} into the expression for the dilute limit of the single inclusive production cross section given in Eq.~\eqref{Dilute_Single_X_sec} and integrating over  transverse coordinates, we can simply write the dilute limit of the single inclusive production cross section as
\beq
\label{S_Inc_X_sec_noneik}
&&
\frac{d\sigma}{d^2kd\eta}\bigg|_{\rm dilute}= 4\pi\, \alpha_s\, C_A\, g^2\int dx_1^+ \, dx_2^+ \int \frac{d^2q_1}{(2\pi)^2}\, \frac{d^2q_2}{(2\pi)^2} \,\delta^{c\bar c} \Big\langle A^-_c(x_1^+,q_1) A^-_{\bar c}(x_2^+, q_2)\Big\rangle_T
\nonumber\\
&&
\hspace{3cm}
\times\; 
\mu^2\big[ k-q_1, q_2-k\big] L^i(k,q_1)L^i(k,q_2),
\eeq
where $L^i(k,q)$ is the strict eikonal Lipatov vertex~\eqref{EikL}.
%
%

At this point, the effects of finite longitudinal width of the target can be implemented in the single inclusive gluon production cross section. Effectively, the implementation of these effects corresponds to two modifications in the cross section given in Eq. \eqref{S_Inc_X_sec_noneik}. The first modification is to replace the eikonal Lipatov vertices by the non-eikonal ones derived in Section \ref{sec:neLip}: 
\beq
L^i(k,q)\to L^i_{\rm NE}(\underline{k}, q; x^+).
\eeq
The non-eikonal Lipatov vertex given in Eq. \eqref{NEikL} takes into account the finite longitudinal width of the target to all orders as discussed in Section \ref{sec:neLip}. The second modification that is needed to account for the finite longitudinal width of the target is adopting a modified expression for the correlator of two target fields. Since the target has finite longitudinal width, the target fields can be located at two different longitudinal positions. Therefore, for the correlator of two target fields, we consider a generalization of the MV model in which the two colour fields are located at different longitudinal  coordinates and are connected via gauge links along the longitudinal axis~\cite{Altinoluk:2015xuy}.  
In that case, the colour field correlator of two fields can be written as
\begin{align}
\label{correlation}
\Big\langle A^-_{c}(x_1^+,q_1) A^-_{\bar{c}}(x_2^+,q_2)\Big\rangle_T = \delta^{c \bar{c}}\;  n(x_1^+) \, \frac{1}{2 \lambda^+} \Theta\Big(\lambda^+-|x_1^+-x_2^+|\Big) (2\pi)^2 \delta^{(2)}(q_1-q_2)\, |a(q_1)|^2,
\end{align}
where $\lambda^+$ is the colour correlation length in the target and much smaller than the total longitudinal width of the target $L^+$. Moreover,  function  $n(x^+)$ defines the one dimensional target density along the longitudinal axis. For simplicity of the calculation, we assume that this function is constant with a finite support, $n(x^+)=n_0$ for $0\leq x^+\leq L^+$ and 0 elsewhere. Finally,  function $a(q)$ that appears in the definition of the two field correlator is the functional form of the potential in momentum space which is usually taken to be a Yukawa type potential in jet quenching calculations \cite{Kovner:2003zj,CasalderreySolana:2007pr,Wiedemann:2000za,Gyulassy:2000er}: 
\begin{align}
|a(q)|^2=\frac{m^2}{\left(q^2+m^2\right)^2},
\label{eq:Deb}
\end{align} 
with $m$  some Debye screening mass or inverse colour correlation length. We would like to emphasise that in the limit of vanishing correlation length $\lambda^+$ together with a constant potential $a(q)$ and a constant longitudinal target density $n(x_1^+)$, the two target field correlator defined in Eq.~\eqref{correlation} reduces to the standard MV model correlator.  

By implementing these two modifications in the single inclusive gluon production cross section and using the expression of the non-eikonal Lipatov vertex given in Eq.~\eqref{NEikL} together with the two field correlator introduced in Eq.~\eqref{correlation}, we can write the non-eikonal generalization of the dilute limit of the single inclusive gluon cross section which accounts for the finite longitudinal width of the target as 
\beq
\frac{d\sigma}{d^2kd\eta}\bigg|_{\rm dilute}^{\rm NE}&=&4\pi\, \alpha_s\, C_A\, (N_c^2-1)\, g^2 \int \frac{d^2q}{(2\pi)^2} \, \mu^2\big[ k-q,q-k\big] \, L^i(k,q)L^i(k,q)\,  \big|a(q)\big|^2 \nonumber\\
&&
\times\, n_0 \frac{1}{2\lambda^+} \int_0^{L^+} dx_1^+\int_{x_1^+-\lambda^+}^{x_1^++\lambda^+} dx_2^+\, e^{i\frac{k^2}{2k^+}(x_1^+-x_2^+)}.
\eeq
In this expression the non-eikonal Lipatov vertex is incorporated via the phase that appears under the longitudinal coordinate integral, the $\theta$-function provides  the limits of the integral in $x_2^+$ and the one dimensional target density along the longitudinal axis is taken to be constant, $n_0$ for $0\leq x_1^+\leq L^+$.  The integrations over the longitudinal coordinates $x_1^+$ and $x_2^+$ can be performed in a straight forward manner and the final result for the dilute limit of the non-eikonal single inclusive gluon production cross section reads
%
%
%
\beq
\label{NE_S_Inc_Final}
\frac{d\sigma}{d^2kd\eta}\bigg|_{\rm dilute}^{\rm NE}&=& 4\pi\, \alpha_s\, C_A\, (N_c^2-1)\, g^2 \, (n_oL^+)\, {\cal G}_1^{\rm NE}(k^-;\lambda^+)
%
\int \frac{d^2q}{(2\pi)^2} \, \mu^2\big[ k-q,q-k\big] \, L^i(k,q)L^i(k,q)\,  \big|a(q)\big|^2 ,
\eeq
where we have used the fact that $\lambda^+\ll L^+$ for the integration over the longitudinal coordinates. Here, ${\cal G}_1^{\rm NE}(k^-,\lambda^+)$ is the function that encodes all the non-eikonal information of the single inclusive gluon production and reads
\beq
\label{G_1}
{\cal G}_1^{\rm NE}(k^-;\lambda^+)= \frac{1}{k^-\lambda^+}\sin(k^-\lambda^+),
\label{eq:g1fun}
\eeq
with $k^-=\frac{k^2}{2k^+}$.
We would like to emphasize that the factor $(n_0L^+)$ in Eq. \eqref{NE_S_Inc_Final} stands for the the number of scattering centres inside the finite longitudinal extend $L^+$ of the target. In the dilute target limit, we only take account one single scattering both in the amplitude and in the complex conjugate amplitude. Therefore, in this limit this factor will be set to one hereafter and we get
\beq
\label{NE_S_Inc_Final_2}
\frac{d\sigma}{d^2kd\eta}\bigg|_{\rm dilute}^{\rm NE}&=& 4\pi\, \alpha_s\, C_A\, (N_c^2-1)\, g^2 \, {\cal G}_1^{\rm NE}(k^-;\lambda^+)
%
\int \frac{d^2q}{(2\pi)^2} \, \mu^2\big[ k-q,q-k\big] \, L^i(k,q)L^i(k,q)\,  \big|a(q)\big|^2 .
\eeq

Eq. \eqref{NE_S_Inc_Final_2} is the final result for the dilute target limit of the non-eikonal single inclusive gluon production cross section. Note that in the limit of vanishing correlation length $\lambda^+$ one can expand the non-eikonal single inclusive production cross section to second order in $(k^-\lambda^+)$ which corresponds to the single inclusive gluon production cross section at next-to-next-to-eikonal accuracy and the result coincides, as announced, with the expression derived in~\cite{Altinoluk:2015xuy}.

\begin{figure}[h!]
	\centering
	\includegraphics[scale=0.8]{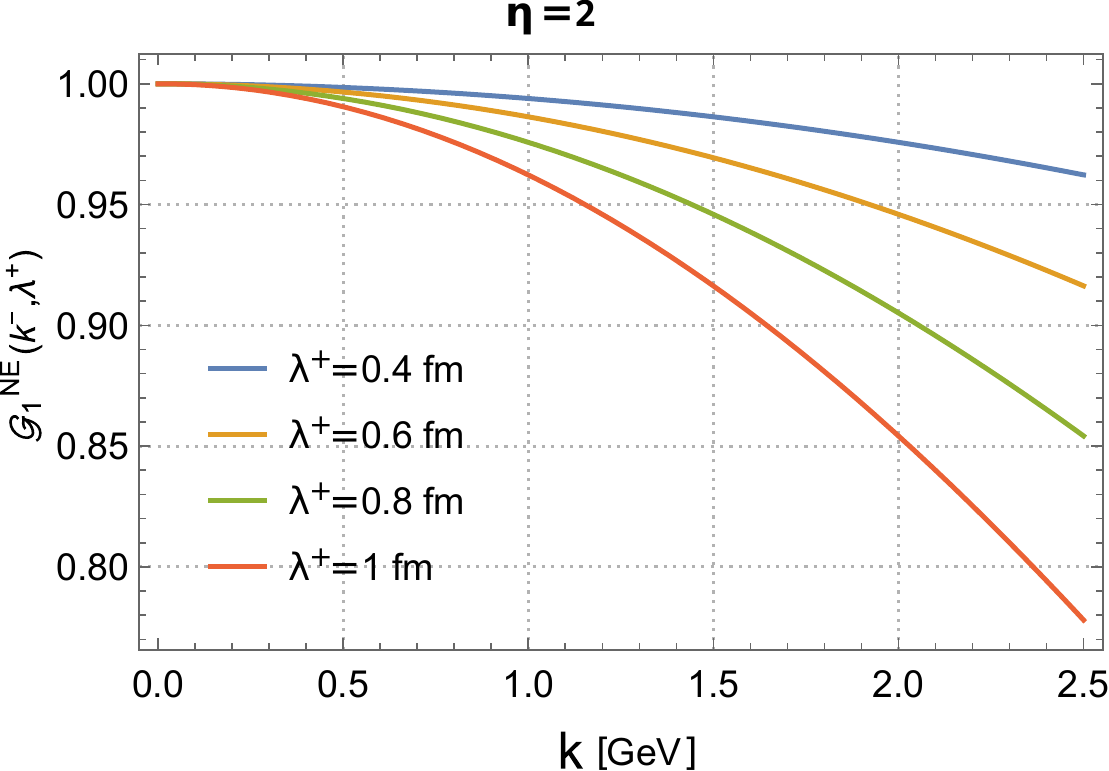}
	\caption{The ratio of non-eikonal to eikonal single inclusive gluon production cross sections,  \eqref{eq:g1fun}, as a function of the transverse momenta of the produced gluon for different values of the correlation length $\lambda^+$, at fixed pseudorapidity $\eta=2$.}
\label{plot_single_1}
\end{figure}

\begin{figure}[h!]
	\centering
	\includegraphics[scale=0.8]{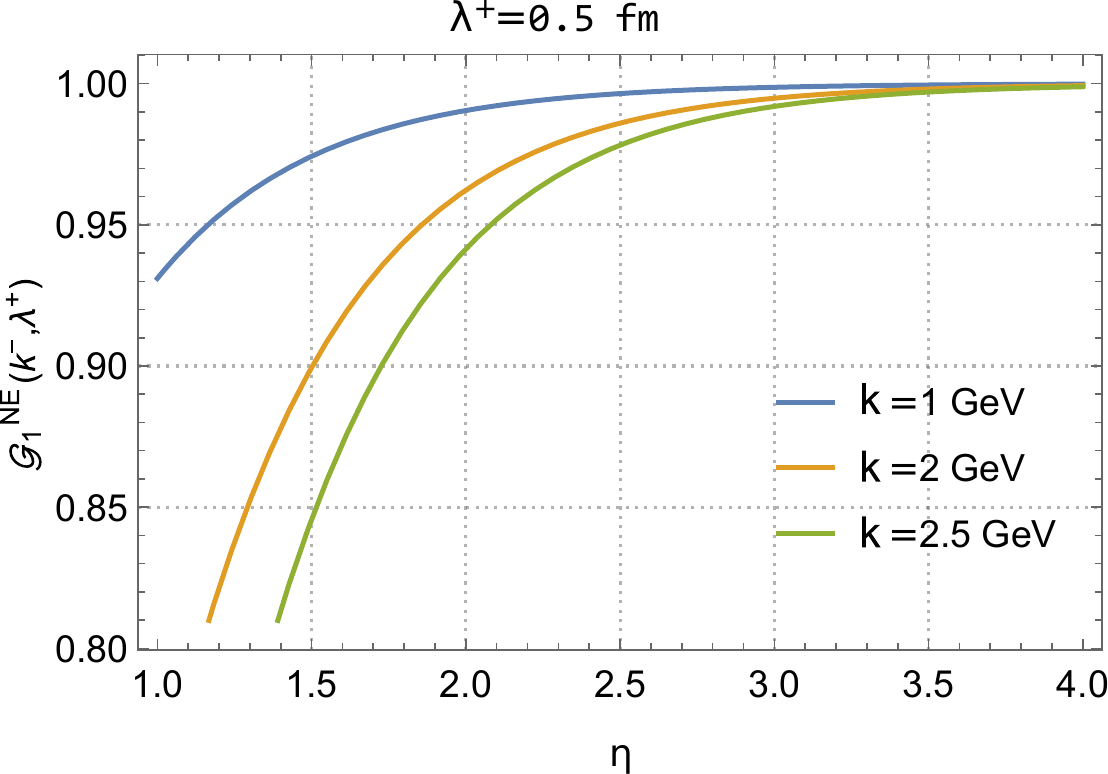}
	\caption{The ratio of non-eikonal to eikonal single inclusive gluon production cross sections,  \eqref{eq:g1fun}, as a function of the pseudorapidity of the produced gluon for different values of its transverse momenta at a fixed correlation length $\lambda^+=0.5$ fm.  }
	\label{plot_single_2}
\end{figure}

Before we conclude this subsection, let us comment on the relative importance of the non-eikonal corrections, that are accounted for  in Eq. \eqref{NE_S_Inc_Final_2} via the function ${\cal G}_1^{\rm NE}(k^-;\lambda^+)$ that encodes the non-eikonal effects, with respect to the eikonal limit of the single inclusive gluon production cross section in the dilute target limit. First of all, in the limit of vanishing $(k^-\lambda^+)$, we have 
\beq
\lim_{k^-\lambda^+\to 0} {\cal G}_1^{\rm NE}(k^-;\lambda^+)=1
\eeq 
and we recover the well known eikonal limit of the single inclusive gluon production in the limit of the dilute target. In Fig.~\ref{plot_single_1}, we have plotted the ratio of the non-eikonal to eikonal single inclusive gluon production cross sections, \eqref{eq:g1fun}, as a function of the transverse momenta of the produced gluon at fixed pseudorapidity $\eta=2$ for different values of the colour correlation length $\lambda^+$. In the limit of vanishing transverse momenta of the produced gluon, the non-eikonal and eikonal cross sections coincide and the ratio becomes one as expected. The ratio shows up to $20\%$ relative weight  of the non-eikonal corrections for $\lambda^+=1$ fm, for smaller values of $\lambda^+$ the results show a suppression from a few to up to $10\%$. 

In Fig.~\ref{plot_single_2}, we have plotted the ratio of the non-eikonal to eikonal single inclusive gluon production cross sections, \eqref{eq:g1fun}, as a function of pseudorapidity for different values of the transverse momenta of the produced gluon at a fixed correlation length $\lambda^+=0.5$ fm. The ratio of the non-eikonal to eikonal cross sections goes to one with increasing pseudorapidity as expected, since the relative importance of the non-eikonal corrections should vanish for large values of $\eta$. The results show that up to pseudorapidity $\eta=2.5$, depending on the value of the transverse momenta of the produced gluon, the relative weight of the non-eikonal corrections can vary roughly between $15\%$ and $2\%$.  These results confirm our analytical predictions for the importance of the non-eikonal corrections in certain kinematical regions.  

\subsection{Double inclusive gluon production beyond the eikonal approximation}
\label{sec:double}

In this Subsection we consider double inclusive gluon production beyond the eikonal approximation. Our strategy for this subsection is the same as the calculation performed for single inclusive gluon production in the previous Subsection. Namely, we start with the double inclusive gluon production cross section that takes into account multiple scatterings in the target in~\cite{Altinoluk:2018ogz}. Then, we consider the dilute target limit of this expression which effectively corresponds to the  Glasma graph approximation by expanding the dipole and quadrupole operators in powers of the background field of the target. Finally, we introduce the finite longitudinal width of the target effects via the non-eikonal Lipatov vertex Eq.~\eqref{NEikL} and the generalised MV model for the two field correlator Eq.~\eqref{correlation} in the expanded expression of the double inclusive gluon production cross section. 

The general expression for the production of two gluons with pseudorapidities $\eta_1$ and $\eta_2$, and with transverse momenta $k_1$ and $k_2$ reads
\beq
\frac{d\sigma}{d^2k_1d\eta_1d^2k_2d\eta_2}&=&\alpha_s^2(4\pi)^2\int_{z_1\bar z_1 z_2\bar z_2} e^{ik_1\cdot(z_1-\bar z_1)+ik_2\cdot(\bar z_2-y_2)} 
\int_{x_1x_2y_1y_2} A^i(x_1-z_1)A^i(\bar z_1-y_1)A^j(x_2-z_2)A^j(\bar z_2-y_2) \nonumber\\
&&
\times\, 
\Big\langle \rho^{a_1}_{x_1}\rho^{a_2}_{x_2}\rho^{b_1}_{y_1}\rho^{b_2}_{y_2}\Big\rangle_P\, 
\Big\langle \big[ U_{z_1}-U_{x_1}\big]^{a_1c} \big[U^\dagger_{\bar z_1}-U^\dagger_{y_1}\big]^{cb_1} \big[ U_{z_2}-U_{x_2}\big]^{a_2d}\big[U^\dagger_{\bar z_2}-U^\dagger_{y_2}\big]^{db_2}\Big\rangle_T.
\eeq 
In the dilute limit of the target, or equivalently in the Glasma graph approximation, the Wilson lines are expanded in powers of the background field of the target as in Eq.~\eqref{expanded_U}. Therefore, in the dilute target limit double inclusive gluon production cross section can be written as 
\beq
\frac{d\sigma}{d^2k_1d\eta_1d^2k_2d\eta_2}\bigg|_{\rm dilute}&=&\alpha_s^2(4\pi)^2\int_{z_1\bar z_1 z_2\bar z_2} e^{ik_1\cdot(z_1-\bar z_1)+ik_2\cdot(\bar z_2-y_2)} 
\nonumber\\
&&
\hspace{-3cm}
\times\, 
\int_{x_1x_2y_1y_2} A^i(x_1-z_1)A^i(\bar z_1-y_1)A^j(x_2-z_2)A^j(\bar z_2-y_2)\, \Big\langle \rho^{a_1}_{x_1}\rho^{a_2}_{x_2}\rho^{b_1}_{y_1}\rho^{b_2}_{y_2}\Big\rangle_P
\nonumber\\
&&
\hspace{-3cm}
\times\, 
g^4\int dx_1^+dx_2^+dx_3^+dx_4^+\int \frac{d^2q_1}{(2\pi)^2} \frac{d^2q_2}{(2\pi)^2} \frac{d^2q_3}{(2\pi)^2} \frac{d^2q_4}{(2\pi)^2}
\Big\langle A^-_a(x_1^+,q_1)A^-_b(x_2^+,q_2)A^-_c(x_3^+,q_3)A^-_d(x_4,q_4)\Big\rangle_T
\nonumber\\
&&
\hspace{-3cm}
\times\, 
(T^aT^b)_{a_1b_1}(T^cT^d)_{a_2b_2} 
\big[ e^{-iq_1\cdot z_1}-e^{-iq_1\cdot x_1}\big]  \big[ e^{iq_2\cdot \bar z_1}-e^{iq_2\cdot y_1}\big]  \big[ e^{-iq_3\cdot z_2}-e^{-iq_3\cdot x_2}\big]  \big[ e^{iq_4\cdot \bar z_2}-e^{iq_4\cdot y_2}\big].
\eeq
Let us now perform the averaging of the double inclusive production cross section with respect to the colour charge densities of the projectile. Since we are using a generalized MV model, the average of any product of the colour charge densities factorize into products of all possible Wick contractions: 
\beq
\label{4rho_Wick}
\Big\langle \rho^{a_1}_{x_1}\rho^{a_2}_{x_2}\rho^{b_1}_{y_1}\rho^{b_2}_{y_2}\Big\rangle_P=
\Big\langle \rho^{a_1}_{x_1}\rho^{a_2}_{x_2}\Big\rangle_P\Big\langle \rho^{b_1}_{y_1}\rho^{b_2}_{y_2}\Big\rangle_P
+
\Big\langle \rho^{a_1}_{x_1}\rho^{b_1}_{y_1}\Big\rangle_P \Big\langle \rho^{a_2}_{x_2}\rho^{b_2}_{y_2}\Big\rangle_P 
+
\Big\langle \rho^{a_1}_{x_1}\rho^{b_2}_{y_2}\Big\rangle_P \Big\langle \rho^{a_2}_{x_2}\rho^{b_1}_{y_1} \Big\rangle_P.
\eeq
For the correlator of two colour charge densities, we use the generalized MV model introduced in Eq.~\eqref{proj_corr}.  After implementing Eq.~\eqref{4rho_Wick}, the dilute limit of the double inclusive gluon production cross section can be written as a sum of three contributions: 
\beq
\label{Double_Inc_all}
&&
\hspace{-0.4cm}
\frac{d\sigma}{d^2k_1d\eta_1d^2k_2d\eta_2}\bigg|_{\rm dilute}=\alpha_s^2(4\pi)^2 g^4 \int_{z_1\bar z_1 z_2\bar z_2} \!\!\!\!\!\!\!e^{ik_1\cdot(z_1-\bar z_1)+ik_2\cdot(\bar z_2-y_2)} 
\int_{x_1x_2y_1y_2} \!\!\!\!\!\!\! A^i(x_1-z_1)A^i(\bar z_1-y_1)A^j(x_2-z_2)A^j(\bar z_2-y_2)\, \
\nonumber\\
&&
\hspace{-0.4cm}
\times\, 
\Big\{ \tr\big[T^aT^bT^dT^c\big]\mu^2(x_1,x_2)\mu^2(y_1,y_2)+ \tr\big[T^aT^b\big]\tr\big[T^cT^d\big]\mu^2(x_1,y_1)\mu^2(x_2,y_2)+\tr\big[T^aT^bT^cT^d\big]\mu^2(x_1,y_2)\mu^2(x_2,y_1)\Big\}
\nonumber\\
&&
\hspace{-0.4cm}
\times\, 
\int dx_1^+dx_2^+dx_3^+dx_4^+\int \frac{d^2q_1}{(2\pi)^2} \frac{d^2q_2}{(2\pi)^2} \frac{d^2q_3}{(2\pi)^2} \frac{d^2q_4}{(2\pi)^2}
\Big\langle A^-_a(x_1^+,q_1)A^-_b(x_2^+,q_2)A^-_c(x_3^+,q_3)A^-_d(x_4,q_4)\Big\rangle_T
\nonumber\\
&&
\hspace{-0.4cm}
\times\, 
\big[ e^{-iq_1\cdot z_1}-e^{-iq_1\cdot x_1}\big]  \big[ e^{iq_2\cdot \bar z_1}-e^{iq_2\cdot y_1}\big]  \big[ e^{-iq_3\cdot z_2}-e^{-iq_3\cdot x_2}\big]  \big[ e^{iq_4\cdot \bar z_2}-e^{iq_4\cdot y_2}\big].
\eeq
In order to preserve the consistency of the notations introduced for different contributions in \cite{Altinoluk:2018ogz}, here after we refer to the first contribution as ${\rm Type\, A}$, the second one as ${\rm Type\, B}$ and the last one as ${\rm Type\, C}$, in Eq.~\eqref{Double_Inc_all}. 

Let us focus on ${\rm Type\,  A}$ contributions to the dilute limit of the double inclusive gluon production cross section and adopt the same procedure applied in single inclusive gluon production in order to incorporate the non-eikonal effects due to the finite longitudinal thickness of the target. The same procedure and arguments hold for the calculation of ${\rm Type\, B}$ and ${\rm Type \, C}$ contributions. After integrating over the transverse coordinates, the ${\rm Type \, A}$ contribution can be written as 
\beq
\frac{d\sigma^{\rm Type \, A}}{d^2k_1d\eta_1d^2k_2d\eta_2}\bigg|_{\rm dilute}&=& \alpha_s^2 (4\pi)^2 \, g^4\,  \tr\big[T^aT^bT^dT^c\big]
\int dx_1^+dx_2^+dx_3^+dx_4^+
\int \frac{d^2q_1}{(2\pi)^2} \frac{d^2q_2}{(2\pi)^2} \frac{d^2q_3}{(2\pi)^2} \frac{d^2q_4}{(2\pi)^2}
\nonumber\\
&&
\hspace{-1cm}
\times\, 
 \Big\langle A^-_a(x_1^+,q_1)A^-_b(x_2^+,q_2)A^-_c(x_3^+,q_3)A^-_d(x_4,q_4)\Big\rangle_T
 \nonumber\\
 &&
\hspace{-1cm}
 \times\, 
 \mu^2\big[k_1-q_1, k_2+q_4\big] \, \mu^2\big[ q_2-k_1, -k_2-q_3\big] 
 L^i(k_1,q_1)L^i(k_1,q_2)\, L^j(k_2,-q_3)L^j(k_2,-q_4).
\eeq
Moreover, we can factorize the the average of the colour fields of the target into all possible Wick contractions and write it in the following factorized way: 
\beq
 \Big\langle A^-_a(x_1^+,q_1)A^-_b(x_2^+,q_2)A^-_c(x_3^+,q_3)A^-_d(x_4,q_4)\Big\rangle_T&=& 
 \Big\langle A^-_a(x_1^+,q_1)A^-_b(x_2^+,q_2) \Big\rangle_T   \Big\langle  A^-_c(x_3^+,q_3)A^-_d(x_4,q_4) \Big\rangle_T 
 \nonumber\\
 &
 +&
 \Big\langle A^-_a(x_1^+,q_1) A^-_d(x_4,q_4)\Big\rangle_T   \Big\langle A^-_c(x_3^+,q_3) A^-_b(x_2^+,q_2) \Big\rangle_T
 \nonumber\\
 &
 +&
 \Big\langle A^-_a(x_1^+,q_1) A^-_c(x_3^+,q_3)\Big\rangle_T   \Big\langle A^-_b(x_2^+,q_2) A^-_d(x_4,q_4) \Big\rangle_T.
\eeq
We can now incorporate the non-eikonal effects due to the finite width of the target. This is achieved by replacing the  Lipatov vertices by the non-eikonal ones and using the generalized MV model for the correlator of two target fields as defined in Eq.~\eqref{correlation}. After implementing these two modifications, the ${\rm Type \, A}$ contribution to the dilute limit of the non-eikonal double inclusive gluon production cross section reads
\beq
&&
\hspace{-0.7cm}
\frac{d\sigma^{\rm Type \, A}}{d^2k_1d\eta_1 \, d^2k_2\eta_2}\bigg|_{\rm dilute}^{\rm NE}=\alpha_s^2\, (4\pi)^2 \, g^4 \, C_A^2\,  (N_c^2-1)
\int \frac{d^2q_1}{(2\pi)^2} \,  \frac{d^2q_2}{(2\pi)^2}\,  \big| a(q_1)\big|^2 \,  \big| a(q_2)\big|^2
\int dx_1^+dx_2^+dx_3^+dx_4^+
\nonumber\\
&&
\hspace{-0.7cm}
\times\; 
e^{ik_1^-(x_1^+-x_2^+)+ik_2^-(x_4^+-x_3^+)} 
\bigg\{
\mu^2\big[ k_1-q_1,k_2-q_2\big]\, \mu^2\big[ q_1-k_1,q_2-k_2\big]\,
L^i(k_1,q_1)L^i(k_1,q_1)\, L^j(k_2,q_2)L^j(k_2,q_2)
\nonumber\\
&&
\hspace{3.4cm}
\times\; 
\frac{1}{2\lambda^+}n(x_1^+)\Theta\Big( \lambda^+-|x_1^+-x_2^+|\Big)\, 
\frac{1}{2\lambda^+}n(x_3^+)\Theta \Big( \lambda^+-|x_3^+-x_4^+|\Big)
\nonumber\\
&&
\hspace{3.4cm}
+\; 
\mu^2\big[ k_1-q_1, k_2+q_1\big]\, \mu^2\big[ q_2-k_1, -k_2-q_2\big] \, 
L^i(k_1,q_1)L^i(k_1,q_2)\, L^j(k_2,-q_2)L^j(k_2,-q_1)
\nonumber\\
&&
\hspace{3.4cm}
\times\; 
\frac{1}{2\lambda^+}n(x_1^+)\Theta\Big( \lambda^+-|x_1^+-x_4^+|\Big)\, 
\frac{1}{2\lambda^+}n(x_2^+)\Theta\Big( \lambda^+-|x_2^+-x_3^+|\Big)
\nonumber\\
&&
\hspace{3.4cm}
+\; 
\frac{1}{2}\mu^2\big[ k_1-q_1, k_2-q_2\big]\, \mu^2\big[ q_2-k_1, q_1-k_2\big] \, 
L^i(k_1,q_1)L^i(k_1,q_2)\, L^j(k_2,q_1)L^j(k_2,q_2)
\nonumber\\
&&
\hspace{3.4cm}
\times\; 
\frac{1}{2\lambda^+}n(x_1^+)\Theta\Big(\lambda^+-|x_1^+-x_3^+|\Big)\, 
\frac{1}{2\lambda^+}n(x_2^+)\Theta\Big(\lambda^+-|x_2^+-x_4^+|\Big) \bigg\},
\eeq
where we have used the following colour identities
\beq
\tr\big[ T^aT^aT^bT^b\big]&=&C_A^2\, (N_c^2-1),\\
\tr\big[ T^aT^bT^aT^b\big]&=&\frac{1}{2}\,C_A^2\, (N_c^2-1) ,
\eeq
with $C_A=N_c$ the quadratic Casimir in the adjoint representation.
Now, the integral over the longitudinal coordinates can be performed in the same way as in the single inclusive gluon production. After using the $\Theta\,$-functions to determine the limits of the integrals, a straightforward integration gives
\beq
\label{Type_A_final}
&&
\frac{d\sigma^{\rm Type \, A}}{d^2k_1d\eta_1 \, d^2k_2\eta_2}\bigg|_{\rm dilute}^{\rm NE}=\alpha_s^2\, (4\pi)^2 \, g^4 \, C_A^2\,  (N_c^2-1)
\int \frac{d^2q_1}{(2\pi)^2} \,  \frac{d^2q_2}{(2\pi)^2}\,  \big| a(q_1)\big|^2 \,  \big| a(q_2)\big|^2
{\cal G}_1^{\rm NE}(k_1^-; \lambda^+)\,{\cal G}_1^{\rm NE}(k_2^-; \lambda^+) 
\nonumber\\
&&
\hspace{1cm}
\times\; 
\bigg\{ 
\mu^2\big[ k_1-q_1,k_2-q_2\big]\, \mu^2\big[q_1-k_1, q_2-k_2\big]\, L^i(k_1,q_1)L^i(k_1,q_1)\, L^j(k_2,q_2)L^j(k_2,q_2)
\\
&&
\hspace{1.5cm}
+\, {\cal G}_2^{\rm NE}(k_1^-,-k_2^-;L^+) \mu^2\big[ k_1-q_1, k_2+q_1\big]\, \mu^2\big[ q_2-k_1,-k_2-q_2\big]\, 
L^i(k_1.q_1)L^i(k_1,q_2)\, L^j(k_2,-q_2)L^j(k_2,-q_1)
\nonumber\\
&&
\hspace{1.5cm}
+\, \frac{1}{2}{\cal G}_2^{\rm NE}(k_1^-,k_2^-; L^+)\,
\mu^2\big[k_1-q_1, k_2-q_2\big]\, \mu^2\big[q_2-k_1, q_1-k_2\big] \, 
L^i(k_1,q_1)L^i(k_2,q_2)\, L^j(k_2,q_1)L^j(k_2,q_2)
\, \bigg\},\nonumber
\eeq
where, on top of the function ${\cal G}_1^{\rm NE}(k^-;\lambda^+)$ that takes into account the non-eikonal effects defined in Eq.~\eqref{G_1}, we have introduced a new function ${\cal G}_2^{\rm NE}(k_1^-,k_2^-; L^+)$ that also accounts for the non-eikonal effects in the dilute target limit of the double inclusive gluon production cross section and reads
\beq
\label{G_2}
{\cal G}_2^{\rm NE}(k_1^-,k_2^-; L^+)= \Bigg\{ \frac{2}{\big(k_1^- -k_2^-\big)L^+}\sin\bigg[\frac{\big(k_1^- -k_2^-\big)}{2}L^+\bigg]\Bigg\}^2.
\label{eq:gNE2}
\eeq
Again, this function goes to 1 when we consider the shockwave (eikonal) limit $L^+ \to 0$.

The same procedure can be adopted to calculate ${\rm Type\, B}$ and ${\rm Type \, C}$ contributions to the dilute target limit of the non-eikonal double inclusive gluon production cross section. The results read
\beq
\label{Type_B_final}
&&
\frac{d\sigma^{\rm Type \, B}}{d^2k_1d\eta_1 \, d^2k_2\eta_2}\bigg|_{\rm dilute}^{\rm NE}=\alpha_s^2\, (4\pi)^2 \, g^4 \, C_A^2\,  (N_c^2-1)
\int \frac{d^2q_1}{(2\pi)^2} \,  \frac{d^2q_2}{(2\pi)^2}\,  \big| a(q_1)\big|^2 \,  \big| a(q_2)\big|^2
{\cal G}_1^{\rm NE}(k_1^-; \lambda^+)\,{\cal G}_1^{\rm NE}(k_2^-; \lambda^+) 
\nonumber\\
&&
\times\, 
\bigg\{
(N_c^2-1)\, \mu^2\big[k_1-q_1,q_1-k_1\big]\, \mu^2\big[ k_2-q_2,q_2-k_2\big]\, L^i(k_1,q_1)L^i(k_1,q_1)\, L^j(k_2,q_2)L^j(k_2,q_2) 
\\
&&
+\; 
{\cal G}_2^{\rm NE}(k_1^-,k_2^-; L^+)\, \mu^2\big[k_1-q_1,q_2-k_1\big]\, \mu^2\big[k_2-q_2,q_1-k_2\big]\, L^i(k_1,q_1)L^i(k_1,q_2)\, L^j(k_2,q_2)L^j(k_2,q_1)
\nonumber\\
&&
+\; 
{\cal G}_2^{\rm NE}(k_1^-,-k_2^-; L^+)\, \mu^2\big[k_1-q_1,q_2-k_1\big]\, \mu^2\big[k_2+q_1,-k_2-q_2\big]\, L^i(k_1,q_1)L^i(k_1,q_2)\, L^j(k_2,-q_1)L^j(k_2,-q_2)
\bigg\}\nonumber
\eeq
and 
\beq
\label{Type_C_final}
&&
\frac{d\sigma^{\rm Type \, C}}{d^2k_1d\eta_1d^2k_2d\eta_2}\bigg|_{\rm dilute}^{\rm NE}=\alpha_s^2\, (4\pi)^2 \, g^4 \, C_A^2\,  (N_c^2-1)
\int \frac{d^2q_1}{(2\pi)^2} \,  \frac{d^2q_2}{(2\pi)^2}\,  \big| a(q_1)\big|^2 \,  \big| a(q_2)\big|^2
{\cal G}_1^{\rm NE}(k_1^-; \lambda^+)\,{\cal G}_1^{\rm NE}(k_2^-; \lambda^+) 
\nonumber\\
&&
\times\, 
\bigg\{
\mu^2\big[ k_1-q_1,q_2-k_2\big]\, \mu^2\big[k_2-q_2,q_1-k_1\big]\, L^i(k_1,q_1)L^i(k_1,q_1)\, L^j(k_2,q_2)L^j(k_2,q_2)
\\
&&
+\; 
{\cal G}_2^{\rm NE}(k_1^-,k_2^-; L^+)\, \mu^2\big[ k_1-q_1, q_1-k_2\big]\, \mu^2\big[k_2-q_2, q_2-k_1\big]\, L^i(k_1,q_1)L^i(k_1,q_2)\, L^j(k_2,q_1)L^j(k_2,q_2)
\nonumber\\
&&
+\; 
\frac{1}{2} {\cal G}_2^{\rm NE}(k_1^-,-k_2^-; L^+)\, \mu^2\big[k_1-q_1,-k_2-q_2\big]\, \mu^2\big[k_2+q_1,q_2-k_1\big]\, L^i(k_1,q_1)L^i(k_1,q_2)\, L^j(k_2,-q_1)L^j(k_2,-q_2)
\bigg\}.\nonumber
\eeq
Finally, we can add the three contributions Eqs. \eqref{Type_A_final}, \eqref{Type_B_final} and \eqref{Type_C_final} and organize the full result of the dilute limit of the non-eikonal double inclusive gluon production cross section as 
\beq
\label{organized_Double_Inc}
\frac{d\sigma}{d^2k_1d\eta_1d^2k_2d\eta_2}\bigg|_{\rm dilute}^{\rm NE}=&&\alpha_s^2\, (4\pi)^2 \, g^4 \, C_A^2\,  (N_c^2-1)
\int \frac{d^2q_1}{(2\pi)^2} \,  \frac{d^2q_2}{(2\pi)^2}\,  \big| a(q_1)\big|^2 \,  \big| a(q_2)\big|^2
{\cal G}_1^{\rm NE}(k_1^-; \lambda^+)\,{\cal G}_1^{\rm NE}(k_2^-; \lambda^+) 
\nonumber\\
&&
\times\, 
\bigg\{ I_{2\tr}^{(0)}+\frac{1}{N_c^2-1}\Big[ I_{2\tr}^{(1)}+I_{1\tr}^{(1)}\Big]\bigg\},
\eeq
where the subscripts denote the single trace terms ($I^{(i)}_{1\tr}$) or the double trace term ($I^{(i)}_{2\tr}$) in the double inclusive gluon production cross section given in Eq.~\eqref{Double_Inc_all}. The explicit expressions for these terms read
\beq
\label{I_2_0}
I_{2\tr}^{(0)}=\mu^2\big[ k_1-q_1,q_1-k_1\big]\, \mu^2\big[ k_2-q_2, q_2-k_2\big]\, L^i(k_1,q_1)L^i(k_1,q_1)\, L^j(k_2,q_2)L^j(k_2,q_2),
\eeq
\beq
\label{I_2_1}
I_{2\tr}^{(1)}=&&\Big\{{\cal G}_2^{\rm NE}(k_1^-,k_2^-; L^+)\mu^2\big[ k_1-q_1,q_2-k_1\big]\, \mu^2\big[k_2-q_2,q_1-k_2\big]
\nonumber\\
&&
\times\; 
 L^i(k_1,q_1)L^i(k_1,q_2)\, L^j(k_2,q_2)L^j(k_2,q_1)\Big\}+(\underline{k}_2\to -\underline{k}_2)
\eeq  
and, finally,
\beq
\label{I_1_1}
&&
I_{1\tr}^{(1)}= \bigg\{
\mu^2\big[k_1-q_1,q_2-k_2\big]\, \mu^2\big[k_2-q_2,q_1-k_1\big] \, L^i(k_1,q_1)L^i(k_1,q_1)\, L^j(k_2,q_2)L^j(k_2,q_2)
\nonumber\\
&&
+\; {\cal G}_2^{\rm NE}(k_1^-,k_2^-; L^+)\Big\lgroup \mu^2\big[k_1-q_1,q_1-k_2\big]\, \mu^2\big[k_2-q_2,q_2-k_1\big]+\frac{1}{2}\mu^2\big[k_1-q_1,k_2-q_2\big]\, \mu^2\big[q_2-k_1,q_1-k_2\big]\Big\rgroup
\nonumber\\
&&
\times\, 
L^i(k_1,q_1)L^i(k_1,q_2)\, L^j(k_2,q_1)L^j(k_2,q_2)\bigg\}+ (\underline{k}_2\to-\underline{k}_2).
\eeq
%

Let us now identify the terms that appear in the dilute target limit of the non-eikonal double inclusive gluon production cross section. For this analysis, we follow the procedure introduced in~\cite{Altinoluk:2018ogz}. The function $\mu^2(k,p)$ can be considered as function of the total transverse momenta and a function of the average transverse momenta: 
\beq
\mu^2(k,p)=T\bigg( \frac{k-p}{2}\bigg)\, F\big[ (k+p)R\big],
\eeq
where  function $T$ can be identified with a transverse momentum dependent distribution of the colour charge densities, and  function $F$ is a soft form factor which is peaked when the argument of the function $F$ vanishes and rapidly decreases when $\big| (k+p)R\big|>1$, with $R$ the radius of the projectile.  In our set up, the transverse momenta $k_1-q_1$ and $k_2-q_2$ are the momenta of the two gluons in the projectile, $k_1$ and $k_2$ are the momenta of the two gluons in the final state and the momenta $q_1$ and $q_2$ are the transverse momenta that are transferred from the target to the projectile during their interaction. In such a set up, the (forward/backward) Bose enhancement of the gluons in the projectile is identified by the form factor that is peaked around $(k_1-q_1)\mp(k_2-q_2)$, the (forward/backward) HBT correlations of the final state gluons are identified by the form factor that is peaked around $k_1\mp k_2$ and finally the (forward/backward) Bose enhancement of the gluons in the target is identified by the form factor that is peaked around $q_1\mp q_2$. We proceed to analyse them all:

\begin{itemize}

\item First of all, it is straightforward to realise that the first term in Eq.~\eqref{organized_Double_Inc}, whose explicit expression is given in Eq.~\eqref{I_2_0}, is nothing but the square of the single inclusive gluon emission probability. Therefore, this term is completely factorised and does not give any contribution to the correlated production. 

\item The second contribution to the non-eikonal double inclusive gluon production cross section is given in Eq.~\eqref{I_2_1}. This term is proportional to 
\beq
\label{F_2_1}
\mu^2\big[ k_1-q_1, q_2-k_1\big] \, \mu^2\big[ k_2-q_2, q_1-k_2\big]= T\bigg[ k_1-\frac{(q_1+q_2)}{2}\bigg]T\bigg[ k_2-\frac{(q_1+q_2)}{2}\bigg]\, F^2\Big(\big|q_1-q_2\big|R\Big).
\eeq
The form factor $F$ in Eq.~\eqref{F_2_1} is strongly peaked when the transverse momenta transferred from the target are very close to each other. Therefore, the term defined in Eq. \eqref{I_2_1} is the term responsible for the Bose enhancement in the target wave function. 

\item Let us now consider the third contribution to the double inclusive gluon production cross section which is defined in Eq.~\eqref{I_1_1}. This contribution consists of three different terms:
\begin{itemize}
\item [(i)] The first term in this contribution is proportional to 
\beq
\mu^2\big[k_1-q_1,q_2-k_2\big]\, \mu^2\big[k_2-q_2,q_1-k_1\big]=T^2\bigg[ \frac{(k_1-q_1)}{2}+\frac{(k_2-q_2)}{2}\bigg]\, F^2\Big[ \big| (k_1-q_1)-(k_2-q_2)\big|R\Big].
\eeq
Since the transverse momenta $k_1-q_1$ and $k_2-q_2$ are the momenta of the two gluons in the projectile wave function and the form factor $F$ is peaked around when the momenta of the two gluons in the projectile wave function are close to each other in this term, it is the Bose enhancement contribution in the projectile wave function.
\item[(ii)] The second term in Eq.~\eqref{I_1_1} is proportional to 
\beq
\mu^2\big[k_1-q_1,q_1-k_2\big]\, \mu^2\big[k_2-q_2,q_2-k_1\big]=T\bigg[ \frac{(k_1+k_2)}{2}-q_1\bigg] T\bigg[ \frac{(k_1+k_2)}{2}-q_2\bigg] F^2\Big[ \big|k_1-k_2\big|R\Big].
\eeq
Now the form factor $F$ is peaked for the transverse momenta of the two gluons in the final state is close to each other, so this term corresponds to the HBT contribution.
\item[(iii)] The last term in Eq.~\eqref{I_1_1} is proportional to 
\beq
\mu^2\big[k_1-q_1,k_2-q_2\big]\, \mu^2\big[q_2-k_1,q_1-k_2\big]&=& T\bigg[ \frac{(k_1-q_1)}{2}-\frac{(k_2-q_2)}{2}\bigg] T\bigg[ \frac{(k_2+q_2)}{2}-\frac{(k_1+q_1)}{2}\bigg] 
\nonumber\\
&
\times&
F^2\Big[ \big| (k_1-q_1)+(k_2-q_2)\big| R\Big].
\eeq
In this term, the form factor is peaked for the transverse momenta of the two gluons in the projectile wave function are close and opposite to each other. Therefore, this term is a contribution to the backward peak of Bose enhancement of gluons in the projectile wave function.
\end{itemize}
\end{itemize}

Apart from the non-eikonal effects that are encoded in the functions ${\cal G}_1^{\rm NE}(k^-;\lambda^+)$ and ${\cal G}_2^{\rm NE}(k_1^-,k_2^-; L^+)$, the main difference between the dilute target limit of the double inclusive gluon production cross section calculated in this subsection and the double inclusive gluon production cross section derived in~\cite{Altinoluk:2018ogz} is the $N_c$ counting of some of the contributions. Our main result, Eq.~\eqref{organized_Double_Inc}, shows that apart from the uncorrelated contribution that is identified as the square of the single inclusive gluon production cross section, all terms that contribute to the correlated production come with the same $N_c$ power. However, in~\cite{Altinoluk:2018ogz}, the Bose enhancement contribution of the gluons in the target and part of the Bose enhancement contribution of the gluons in the projectile have shown to be $N_c$-suppressed  with respect to the rest of the terms. This is a well known consequence of the fact that some aspects of $N_c$ counting are different in the dilute and dense limits~\cite{Altinoluk:2014mta,Altinoluk:2014twa}.

Let us comment on the function ${\cal G}_2^{\rm NE}(k_1^-,k_2^-; L^+)$, Eq.~\eqref{eq:gNE2}, which is one of the functions that encode the non-eikonal effects in the double inclusive gluon production in the dilute target limit. As it can be seen clearly from the final expression, Eq.~\eqref{organized_Double_Inc} together with Eqs.~\eqref{I_2_0}, \eqref{I_2_1} and \eqref{I_1_1}, the mirror image of the terms that  contribute to the correlated production of two gluons which is given by $(\underline k_2\to -\underline k_2)$, is accompanied by ${\cal G}_2^{\rm NE}(k_1^-,-k_2^-; L^+)$.  However, in certain kinematic regimes the behaviour of ${\cal G}_2^{\rm NE}(k_1^-,k_2^-; L^+)$ differs completely from ${\cal G}_2^{\rm NE}(k_1^-,-k_2^-; L^+)$. Namely, in the kinematic region where  $k_1^-\sim k_2^-$ we get 
\beq
\label{behaviour_G2}
{\cal G}_2^{\rm NE}(k_1^-,k_2^-; L^+)\gg {\cal G}_2^{\rm NE}(k_1^-,-k_2^-; L^+)
\eeq
which creates an asymmetry between the terms with $(\underline k_1,\underline k_2)$ and their partners with $(\underline k_2\to -\underline k_2)$. This asymmetry created by the  non-eikonal effects immediately reminds  the asymmetry between the forward and backward peaks of the ridge structure observed in two particle production.

While a dedicated study of two particle correlations and azimuthal harmonics with non-eikonal corrections is left for a forthcoming work \cite{Azimuthal_Harm_NonEik}, here we show a few results with the sole purpose of illustratining these points. To compute them, we have taken $N_c=3$, $m=0.2$ GeV in~\eqref{eq:Deb}, $\mu^2(k,q)\propto \delta^{(2)}(k+q)$ (i.e. translational invariance) but with a projectile size $S_\perp=4$ GeV$^{-2}$, and regulate the denominators that give rise to infrared divergencies by substituting the corresponding squared transverse momenta $l^2 \to l^2+m_g^2$ where we have used the numerical value $m_g^2=0.2$ GeV.

\begin{figure}[h!]
	\centering
	\includegraphics[scale=0.8]{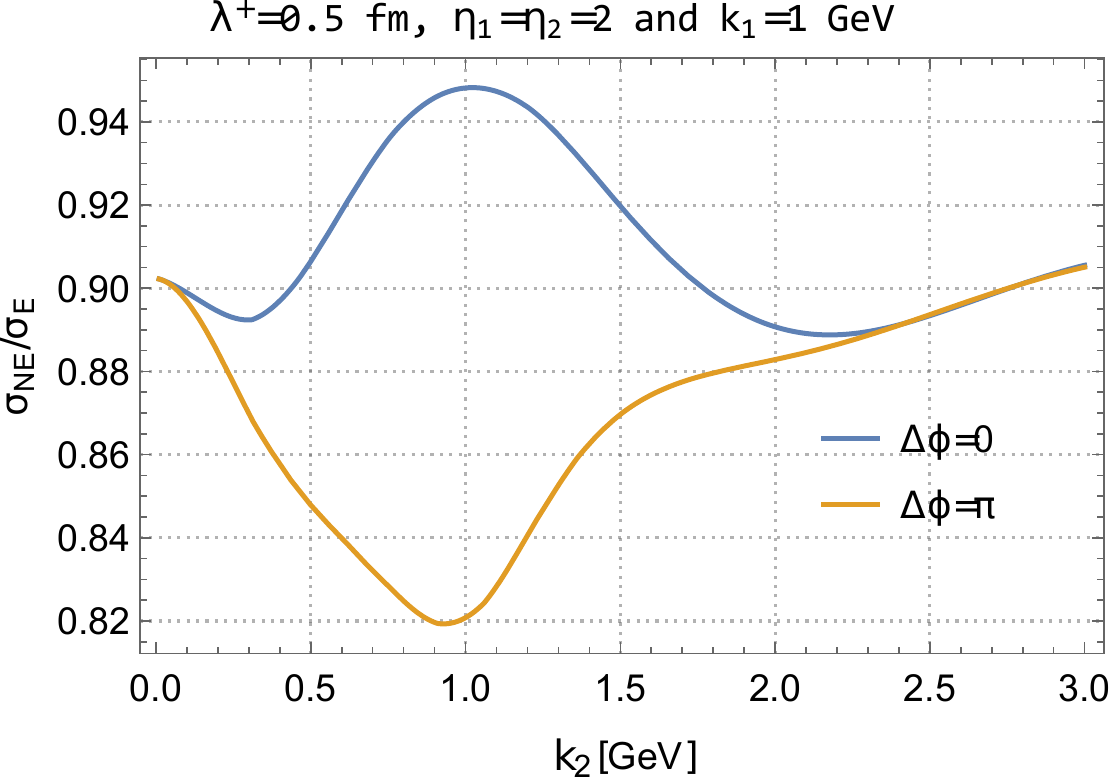}
	\caption{The behaviour of the ratio of non-eikonal to eikonal cross sections at $\Delta\phi=0$ and $\Delta\phi=\pi$ as a function of the transverse momenta of the second gluon for a correlation length $\lambda^+=0.5$ fm, $L^+=6$ fm,  rapidities of the produced gluons $\eta_1=\eta_2=2$ and  transverse momenta of the first gluon $k_1=1$ GeV. }
	\label{double_inc_1}
\end{figure}

In Fig.~\ref{double_inc_1} we show the ratio of the non-eikonal to eikonal double inclusive gluon production cross sections as a function of the transverse momenta of the second produced gluon  while keeping the transverse momenta of the first gluon fixed $k_1=1$ GeV, for $\Delta\phi=0$ and $\Delta\phi=\pi$ with $\Delta \phi$ the azimuthal angle between the two produced gluons. In this plot, we use for the correlation length $\lambda^+=0.5$ fm, $L^+=6$ fm and the pseudorapidities of the produced gluons $\eta_1=\eta_2=2$. The result shows that the ratio of the non-eikonal and eikonal double inclusive gluon cross sections is enhanced for $\Delta\phi=0$ and suppressed for $\Delta \phi=\pi$ as expected by our observation for the behaviour of  ${\cal G}_2^{\rm NE}(k_1^-,k_2^-; L^+)$ given in Eq.~\eqref{behaviour_G2}. The relative modification is peaked when the transverse momenta of the second gluon is  the same as the transverse momenta of the first gluon and it varies roughly between $4\%$ and $10\%$ for  values of the transverse momenta of the second gluon $0.5$ GeV $< k_2 <1.5$ GeV.

\begin{figure}[h!]
	\centering
	\includegraphics[scale=0.8]{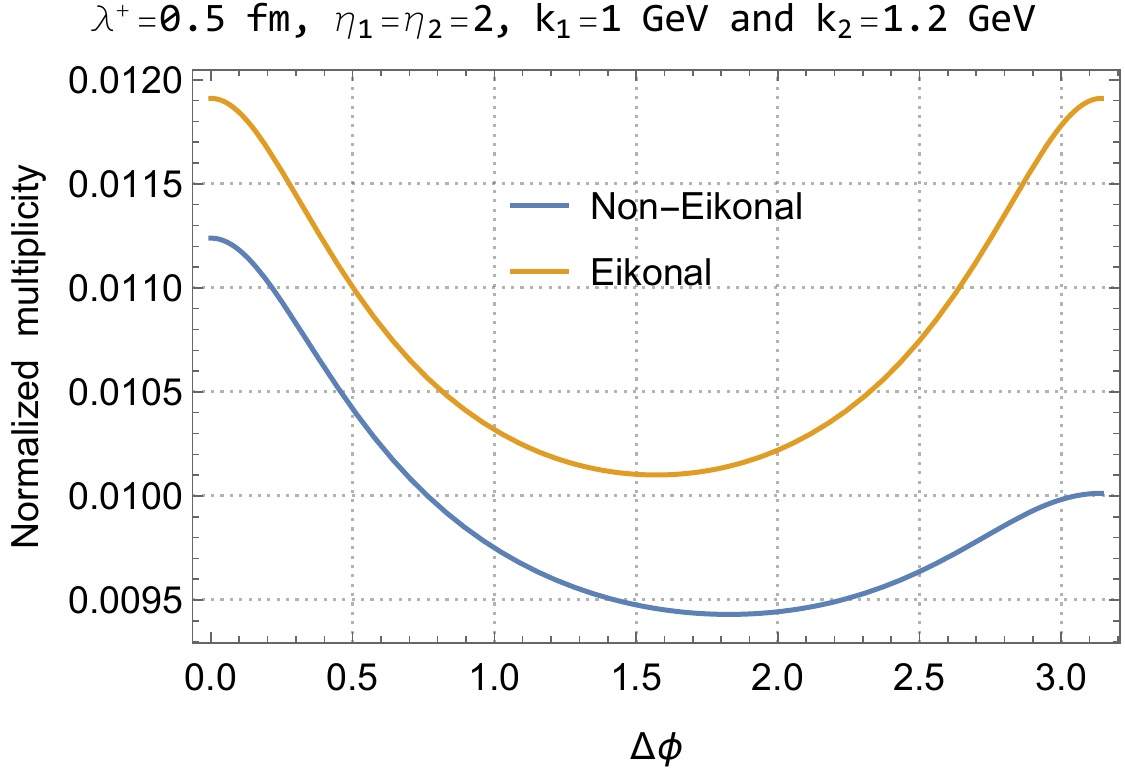}
	\caption{The non-eikonal and eikonal normalized double inclusive gluon production cross sections as a function of azimuthal angle between the two produced gluons $\Delta\phi$ for  $\lambda^+=0.5$ fm, $L^+=6$ fm, and rapidities $\eta_1=\eta_2=2$ and  transverse momenta $k_1=1$ GeV and $k_2=1.2$ GeV of the two produced gluons.  }
	\label{double_inc_2}
\end{figure}

In Fig.~\ref{double_inc_2} we  plot the normalized  non-eikonal and eikonal double inclusive gluon production cross sections as a function of the azimuthal angle between the two produced gluons $\Delta\phi$. We again take $\lambda^+=0.5$ fm, $L^+=6$ fm, the rapidities of the two produced gluons $\eta_1=\eta_2=2$ and their transverse momenta $k_1=1$ GeV and $k_2=1.2$ GeV. These kinematic values are chosen to enhance the asymmetry coming from the behaviour of  function ${\cal G}_2^{\rm NE}(k_1^-,k_2^-; L^+)$. The results are completely symmetric with respect to $\Delta\phi=\pi/2$ in the eikonal case, while an asymmetric behaviour is seen for the non-eikonal case. 

\subsection{Triple inclusive gluon production beyond the eikonal approximation}
\label{sec:triple}

Let us now proceed with the triple inclusive gluon production cross section. The general expression for the production of three gluons, with transverse momenta $k_1$, $k_2$ and $k_3$ and with pseudorapidities $\eta_1$, $\eta_2$ and $\eta_3$ in the dilute-dense set up reads~\cite{Altinoluk:2018ogz}
\begin{align}
&\frac{d \sigma}{d^2 k_1 d \eta_1 d^2 k_2 d \eta_2 d^2 k_3 d \eta_3}=\alpha_s^3 (4\pi)^3 \int_{z_1 z_2 z_3 \bar{z}_1 \bar{z}_2 \bar{z}_3} e^{ik_1\cdot (z_1-\bar{z}_1)+ik_2\cdot (z_2-\bar{z}_2)+ik_3\cdot(z_3-\bar{z}_3)} \nonumber \\
&\times \int_{x_1 x_2 x_3 y_1 y_2 y_3} A^i(x_1-z_1) A^i(\bar{z}_1-y_1) A^j(x_2-z_2) A^j(\bar{z}_2-y_2) A^k(x_3-z_3)A^k(\bar{z}_3-y_3) \Braket{\rho^{a_1}_{x_1}\rho^{a_2}_{x_2}\rho^{a_3}_{x_3}\rho^{b_1}_{y_1}\rho^{b_2}_{y_2}\rho^{b_3}_{y_3}}_P  \nonumber \\
&\times  \Braket{\left\{ \Big[  U_{z_1}-U_{x_1} \Big] \left[  U^\dagger_{\bar{z}_1}-U^\dagger_{y_1} \right] \right\}^{a_1 b_1} \left\{ \Big[  U_{z_2}-U_{x_2} \Big] \left[  U^\dagger_{\bar{z}_2}-U^\dagger_{y_2} \right] \right\}^{a_2 b_2} \left\{ \Big[  U_{z_3}-U_{x_3} \Big] \left[  U^\dagger_{\bar{z}_3}-U^\dagger_{y_3} \right] \right\}^{a_3 b_3}}_T.
\end{align}
After the manipulations described in Appendix \ref{app:triple}, we can organize the dilute target limit of the non-eikonal triple inclusive gluon production cross section according to the powers in the number of colors and the result reads
\begin{align}
&
\frac{d \sigma}{d^2 k_1 d \eta_1 d^2 k_2 d \eta_2 d^2 k_3 d \eta_3}\bigg|_{\rm dilute}^{\rm NE} = (4\pi)^3 \, \alpha_s^3 \, g^6 \, C_A^3\,  (N_c^2-1)^3 \, 
 \int  \frac{d^2q_1}{(2\pi)^2}  \frac{d^2q_2}{(2\pi)^2}  \frac{d^2q_3}{(2\pi)^2} \, \big| a(q_1)\big|^2 \, \big|a(q_2)\big|^2 \, \big|a(q_3)\big|^2 
 \nonumber\\
 &
 \times
 {\cal G}_1(k_1^-;\lambda^+)\, {\cal G}_1(k_2^-;\lambda^+) {\cal G}_1(k_3^-; \lambda^+)  \; 
 \bigg\{  I_{\rm 3tr}^{(0)}+\frac{1}{(N_c^2-1)} \left[I_{\rm 3tr}^{(1)} + I_{\rm 2tr,1}^{(1)}+I_{\rm 2tr,2}^{(1)}\right] 
\nonumber \\
&+
\frac{1}{(N_c^2-1)^2} \Big [ \Big( I_{\rm 3tr,1}^{(2)}+I_{\rm 3tr,2}^{(2)}\Big) +\Big( I_{\rm 2tr,1}^{(2)}+I_{\rm 2tr,2}^{(2)} +I_{\rm 2tr,3}^{(2)} \Big)+ \Big(  I_{\rm 1tr,1}^{(2)}+I_{\rm 1tr,2}^{(2)}+I_{\rm 1tr,3}^{(2)}+I_{\rm 1tr,4}^{(2)} \Big)\Big] \bigg\},
\label{full_triple_prod}
\end{align}
where functions $I_{i{\rm tr},j}^{(k)}$ can be found in Eqs. \eqref{eq:ref0} to \eqref{tilde_3-tr-2-2}, \eqref{2-tr-1-1} to \eqref{tilde_2-tr-3-2} and \eqref{1tr-1-2} to \eqref{tilde_1-tr-4-2_P}.
This is our final result for the dilute target limit of the non-eikonal triple inclusive gluon production cross section. Apart from the fact that this result accounts for the finite longitudinal width target effects through non-eikonal Lipatov vertices which leave their imprints in the functions ${\cal G}_i^{\rm NE}$ upon integration over the longitudinal coordinates, it is valid to all orders in the number of colors. It differs from the dilute target limit of the result calculated in~\cite{Altinoluk:2018ogz} in  two aspects. First, the study performed in~\cite{Altinoluk:2018ogz}, while it is valid for the dense target limit, is truncated at ${\cal O}\big(1/(N_c^2-1)^3\big)$. This obviously affects the total number of terms in the final result. Second, as we will discuss next, some of the $N_c$-suppressed terms that were discarded in~\cite{Altinoluk:2018ogz}, have been identified in our study and shown to establish some interference effects that were absent there.

Let us now consider each term in Eq.~\eqref{full_triple_prod} separately and identify their correlation features. For this analysis we follow the same strategy introduced in Subsection~\ref{sec:double} and use the fact that 
\beq
\mu^2(k,p)\propto F\big[ |k+p|R\big],
\eeq
with $R$ being the radius of the projectile and the form factor $F$ peaked at zero.

\begin{itemize}
\item We start our analysis with the ${\cal O}(1)$ terms. The only ${\cal O}(1)$ term in the dilute target limit of the non-eikonal triple inclusive gluon production cross section is $I_{\rm 3tr}^{(0)}$ term. It is equal to product of three single inclusive gluon production cross sections and it gives contribution to the totally uncorrelated production of three gluons. 

\item Next, we consider the ${\cal O}\big[ 1/(N_c^2-1)\big]$ terms. At this order, we have three different terms: one originating from three-trace contribution and two originating from double-trace contribution.
\begin{itemize}
\item[(i)] The explicit expression of the three-trace term, $I_{\rm 3tr}^{(1)}$, is given in Eq.~\eqref{tilde_3-tr-1-1} and its symmetry partners in Eq.~\eqref{3-tr-1-1}. This term is proportional to 
\beq
\mu^2\big[ k_1-q_1,q_2-k_1\big]\,  \mu^2\big[k_2-q_2,q_1-k_2\big] \,  \mu^2\big[k_3-q_3,q_3-k_3\big] \propto F^2\big[ |q_2-q_1|R\big]  \mu^2\big[k_3-q_3,q_3-k_3\big] 
\eeq
which is clearly a contribution to the forward Bose enhancement of the gluons $q_1$ and $q_2$ while the third gluon is emitted independently from the others. Its mirror image, given by $(\underline k_2\to -\underline k_2)$, exhibits the same behaviour. The symmetry partners of this term which are obtained through $(\underline k_1\leftrightarrow \underline k_3)$ and $(\underline k_2\leftrightarrow \underline k_3)$ correspond to the two cases where the independently emitted gluon is the first and the second gluons, and the remaining two still give contribution to the forward Bose enhancement in the target wave function. 

\item [(ii)] The remaining two terms at this order, originate from the double-trace contribution. The explicit expression of the first of these terms is given in Eq.~\eqref{tilde_2-tr-1-1} and its symmetry partners are given in Eq.~\eqref{2-tr-1-1}.  This term is proportional to 
\beq
&&
 \mu^2\big[ k_1-q_1,q_1-k_1\big]\,  \mu^2\big[ k_2-q_2,q_3-k_3\big] \,  \mu^2\big[k_3-q_3,q_2-k_2\big]\\
 & \propto& F^2\big[ |(k_2-q_2)-(k_3-q_3)| R\big]\,  \mu^2\big[ k_1-q_1,q_1-k_1\big]\nonumber
\eeq
which can be easily identified as a contribution to the forward Bose enhancement of the gluons $k_2-q_2$ and $k_3-q_3$ in the projectile wave function while the first gluon is emitted independently of the remaining two. Clearly, the symmetry partners of this term corresponds to the independent emission of second and third gluons, while the remaining two gives contribution to the forward Bose enhancement in the projectile wave function.  

\item [(iii)] The explicit expression of the last term at this order, $I_{\rm 2tr,2}^{(1)}$, is given in Eq.~\eqref{tilde_2-tr-2-1} with its symmetry partners given in Eq.~\eqref{2-tr-2-1}. This term is proportional to 
\beq
&&
\mu^2\big[ k_3-q_3,q_3-k_3\big]\bigg\{ 
\mu^2\big[ k_1-q_1,q_1-k_2\big] \,  \mu^2\big[ k_2-q_2,q_2-k_1\big] \nonumber \\
&& \hskip 3.3cm +\frac{1}{2} \mu^2\big[k_1-q_1,k_2-q_2\big]\,  \mu^2\big[ q_2-k_1,q_1-k_2\big] \bigg\}
\nonumber\\
&
\propto&
\mu^2\big[ k_3-q_3,q_3-k_3\big]\bigg\{ F^2\big[ |k_1- k_2|R\big] +\frac{1}{2} F^2\big[ |(k_1-q_1)+(k_2-q_2)|R\big]\bigg\}.
\eeq
The first term in the brackets corresponds to forward HBT of the gluons $k_1$ and $k_2$, and the second term corresponds to backward Bose enhancement of the gluons $k_1-q_1$ and $k_2-q_2$ in the projectile wave function while the third gluon is emitted independently from the other two. The mirror image of this term which obtained through $(\underline k_2\to -\underline k_2)$ corresponds to  backward HBT of the gluons $k_1$ and $k_2$, and forward Bose enhancement of the gluons $k_1-q_1$ and $k_2-q_2$ in the projectile wave function while the third gluon is emitted independently.  The symmetry partners of this term which are obtained via $(\underline k_1\leftrightarrow \underline k_3)$ and $(\underline k_2\leftrightarrow \underline k_3)$ correspond to the following two cases:  emission of the first gluon (or the second gluon in the second symmetry partner) while the remaining two gluons exhibit the same behaviour and contribute to (forward/backward) HBT and (backward/forward) projectile Bose enhancement of the corresponding gluons.

\end{itemize}

\item We can now proceed with the ${\cal O}\big[ 1/(N_c^2-1)^2\big]$ terms. At this order, we have terms originating from the  three-trace, the double-trace and the single-trace contributions.
\begin{itemize}
\item [(i)] Let us start with the terms originating from the three-trace contribution:
\begin{itemize}
\item [(a)] The explicit expression for the first term in there, $I_{\rm 3tr,1}^{(2)}$, is given in Eq.~\eqref{tilde_3-tr-1-2} and its symmetry partners are given in Eq.~\eqref{3-tr-1-2}. This term is proportional to 
\beq
&&
\mu^2\big[k_1-q_1,q_2-k_1\big] \, \mu^2\big[ k_2+q_1,q_3-k_2\big]\,  \mu^2\big[k_3-q_2,-q_3-k_3\big]  \\
&\propto& F\big[ |q_2-q_1|R\big] \, F\big[ |q_1+q_3|R \big] F\big[ |q_2+q_3|R \big]. \nonumber
\eeq
This term gives contribution to the case where all three gluons are correlated. In particular, it contributes to forward target Bose enhancement of the gluons $q_1$ and $q_2$ with contributions to backward target Bose enhancement between the gluons $q_1$ and $q_3$ as well as  $q_2$ and $q_3$. Since the form factors in this term are independent of the momenta of the produced gluons, the mirror image of this term and its symmetric partners exhibit exactly the same behaviour.
\item [(b)] The second term in the three-trace contribution at ${\cal O}\big[ 1/(N_c^2-1)^2\big]$ is $I_{\rm 3tr,2}^{(2)}$ which is defined in Eq.~\eqref{tilde_3-tr-2-2} and its symmetric partner is defined in Eq.~\eqref{3-tr-2-2}. This term is proportional to 
\beq
&&
\mu^2\big[k_1-q_1,q_2-k_1\big] \, \mu^2\big[ k_2-q_3,q_1-k_2\big]\,  \mu^2\big[k_3-q_2,q_3-k_3\big] \\
&\propto& F\big[ |q_1-q_2|R\big] \, F\big[ |q_1-q_3|R\big]\, F\big[ |q_2-q_3|R\big].\nonumber
\eeq
Clearly, this term is a contribution to the forward  Bose enhancement of the target gluons between the gluons $q_1$ and $q_2$, together with $q_1$ and $q_3$, as well as $q_2$ and $q_3$. Its symmetric partner defined in Eq.~\eqref{3-tr-2-2} exhibits the same behaviour.

\end{itemize}

Before we continue our analysis with the terms originating from the double-trace contributions at ${\cal O}\big[ 1/(N_c^2-1)^2\big]$, we would like to mention that the two terms $I_{\rm 3tr,1}^{(2)}$ and $I_{\rm 3tr,2}^{(2)}$ give contribution to the correlated production of all three gluons. However, the study performed in~\cite{Altinoluk:2018ogz} has shown that the totally correlated production of three gluons originate from the sextuple contribution which in our case corresponds to the single-trace contribution. This difference is due to the fact that the analogue of the terms $I_{\rm 3tr,1}^{(2)}$ and $I_{\rm 3tr,2}^{(2)}$ in the dense target limit are suppressed in powers of the number of colors and therefore discarded in~\cite{Altinoluk:2018ogz}. In our study, we show that these terms are of the same order as the single-trace contribution and give contribution to the totally correlated production. The difference between the counting of the number of colors in the dilute and dense limits is addressed in detail in~\cite{Altinoluk:2014mta,Altinoluk:2014twa}.

\item [(ii)] Let us proceed with the terms that originate from double-trace contribution at order  ${\cal O}\big[ 1/(N_c^2-1)^2\big]$:
\begin{itemize}
\item[(a)] The first term is $I_{\rm 2tr,1}^{(2)}$ and it is defined in Eq.~\eqref{tilde_2-tr-1-2} with its symmetric partners defined in Eq. \eqref{2-tr-1-2}.  This term is proportional to 
\beq
\label{mu_2-tr-1-2}
&&
\mu^2\big[ k_1-q_1,q_2-k_1\big]
\Big\{ 
\mu^2\big[ k_2-q_2,q_3-k_3\big] \, \mu^2\big[ k_3-q_2,q_1-k_2\big]  \nonumber\\
&&\hskip 3.2cm + \mu^2\big[ k_2-q_2,k_3-q_3\big]\, \mu^2\big[ q_1-k_2,q_3-k_3\big] \Big\}
\\
&
\propto&
F\big[ |q_1-q_2|R\big] \Big\{ F\big[ |(k_2-q_2)-(k_3-q_3)|R\big]\, F\big[ |k_3-k_2|R\big]\,+\, F^2\big[ |(k_2-q_2)+(k_3-q_3)|R\big]\Big\}.\nonumber
\eeq
The first term in Eq.~\eqref{mu_2-tr-1-2} is a contribution to the forward target Bose enhancement of the gluons $q_1$ and $q_2$, together with the forward projectile Bose enhancement of the gluons $k_2-q_2$ and $k_3-q_3$ and forward HBT contribution to the gluons $k_2$ and $k_3$. However, due to the HBT contribution to the gluons $k_2$ and $k_3$, the second form factor in this term can be considered as peaking around $(q_3-q_2)$   and, in that case, it would contribute to the forward Bose enhancement of the gluons $q_2$ and $q_3$ in the gluon wave function. In~\cite{Altinoluk:2018ogz} there were no such contributions, again due to the fact that this term is suppressed in powers of the number of colors in the dense target limit. We would like to mention that, in the translationally invariant limit, this term is suppressed by a phase space integration with respect to the other terms at ${\cal O}\big[ 1/(N_c^2-1)^2\big]$.  The second term in~\eqref{mu_2-tr-1-2} is a contribution to the forward Bose enhancement of the gluons $q_1$ and $q_2$ in the target wave function together with backward contribution the Bose enhancement of the gluons $k_2-q_2$ and $k_3-q_3$ in the projectile wave function. 

\item[(b)] The second term that originates from the double-trace operator at ${\cal O}\big[ 1/(N_c^2-1)^2\big]$ is $I^{(2)}_{\rm 2tr,2}$. It is defined in Eq.~\eqref{tilde_2-tr-2-2} with its symmetry partners in Eq.~\eqref{2-tr-2-2}. This terms has three different pieces. The first piece is proportional to 
\beq
&&
 \mu^2\big[ k_1-q_1,-q_2-k_1\big]  
\bigg\lgroup  \frac{1}{2} \mu^2\big[ k_2+q_1,q_3-k_3\big]  \mu^2\big [k_3+q_2,-q_3-k_2\big]  \nonumber \\
&&\hskip 3.5cm+   \mu^2\big[ k_2+q_1,k_3+q_2\big]  \mu^2\big[ -q_3-k_2,q_3-k_3\big] \bigg\rgroup 
\nonumber\\
&
\propto&
F\big[ |q_1+q_2|R\big]\bigg\lgroup \frac{1}{2}F^2\big[ |(k_2-q_2)-(k_3-q_3)|R\big] \, +\, F^2\big[ |k_2+k_3|R\big]\bigg\rgroup.
\eeq
Clearly, the first term is a contribution to the backward Bose enhancement of the gluons $q_1$ and $q_2$ in the target wave function with a contribution to the forward Bose enhancement of the gluons $k_2-q_2$ and $k_3-q_3$ in the projectile wave function. The second term is a contribution to the backward Bose enhancement of the gluons $q_1$ and $q_2$ in the target wave function with a backward HBT to contribution to the gluons $k_2$ and $k_3$. The second piece of $I^{(2)}_{\rm 2tr,2}$ is proportional to 
\beq
\label{mu_2-tr-2-2}
&&
\mu^2\big[ k_2+q_1,q_2-k_2\big] 
\bigg\lgroup \mu^2\big[ k_1-q_1,-k_3-q_2\big]  \mu^2\big[ q_3-k_1,k_3-q_3\big]    \nonumber \\
&& \hskip 3.3cm+   \frac{1}{2} \mu^2\big[ k_1-q_1,k_3-q_3\big]   \mu^2\big[ q_3-k_1,-k_3-q_2\big] \bigg\rgroup
\nonumber\\
&
\propto &
F\big[ |q_1+q_2|R\big] \bigg\lgroup F^2\big[ |k_1-k_3|R\big] +\frac{1}{2}F^2\big[ |(k_1-q_1)+(k_3-q_3)|R\big] \bigg\rgroup.
\eeq
The first term in Eq.~\eqref{mu_2-tr-2-2} is a contribution to the backward Bose enhancement of the gluons $q_1$ and $q_2$ in the target wave function  with a forward contribution to the HBT of the gluons $k_1$ and $k_3$. The second term in Eq.~\eqref{mu_2-tr-2-2} is a contribution to the backward Bose enhancement of the gluons $q_1$ and $q_2$ in the target wave function with a backward contribution to the Bose enhancement of the gluons $k_1-q_1$ and $k_3-q_3$ in the projectile wave function. The last piece of  $I^{(2)}_{\rm 2tr,2}$ is proportional to 
\beq
\label{mu_2-tr-3-2}
&&
\mu^2\big[ k_3+q_2,-k_3-q_3\big] 
\bigg\lgroup  \frac{1}{2} \mu^2\big[ k_1-q_1,q_3-k_2\big]  \mu^2\big[ -q_2-k_1,k_2+q_1\big]  \nonumber \\
&& \hskip 3.6cm +  \mu^2\big[ k_1-q_1,k_2+q_1\big]  \mu^2\big[- q_2-k_1,q_3-k_2\big] \bigg\rgroup
\\
&
\propto&
F\big[ |q_2-q_3|R\big] \bigg\lgroup \frac{1}{2} F\big[ |(k_1-q_1)-(k_3-q_3)|R\big] F\big[ |(k_1-q_1)-(k_2-q_2)|R\big] \, + \, F^2\big[ |k_1-k_2|R\big]\bigg\rgroup. \nonumber 
\eeq
The first term in this equation is a contribution to forward Bose enhancement of the gluons $q_2$ and $q_3$ in the target wave function together with forward Bose enhancement of the gluons $k_1-q_1$ and $k_3-q_3$ as well as $k_1-q_1$ and $k_2-q_2$ in the projectile wave function. The second term  is a contribution to forward Bose enhancement of the gluons $q_2$ and $q_3$ in the target wave function together with the forward HBT of the gluons $k_1$ and $k_2$. The symmetry partners of all three pieces in $I_{\rm 2tr,2}$ that are defined in Eq.~\eqref{2-tr-2-2} can be easily identified in the same way. 

\item[(c)] The last term that originates from the double-trace contribution is $I_{\rm 2tr,3}^{(2)}$ which is defined in Eq.~\eqref{tilde_2-tr-3-2} together with its symmetry partner defined in Eq.~\eqref{2-tr-3-2}. The first piece in $I_{\rm 2tr,3}^{(2)}$ is proportional to 
\beq
\label{mu_2tr-3-1-2}
&&
\mu^2\big[ k_1-q_1,q_2-k_1\big]  
\bigg\lgroup  \mu^2\big[ k_2-q_3,q_3-k_3\big]  \mu^2\big[ k_3-q_2,q_1-k_2\big] \nonumber \\
&& \hskip 3.3cm +   \frac{1}{2}  \mu^2\big[ k_2-q_3,k_3-q_2\big]  \mu^2\big[ q_1-k_2,q_3-k_3\big]  \bigg\rgroup
\nonumber\\
&
\propto&
F\big[ |q_1-q_2|R\big] \bigg\lgroup F^2\big[ |k_2-k_3|R\big]\, +\,\frac{1}{2} F^2\big[ |(k_2-q_2)+(k_3-q_3)|R\big]\bigg\rgroup.
\eeq
The first term here is clearly a contribution to the forward Bose enhancement of the gluons $q_1$ and $q_2$ in the target wave function together with a contribution to forward HBT of gluons $k_2$ and $k_3$.  The second term is a contribution to the forward Bose enhancement of the gluons $q_1$ and $q_2$ in the target wave function together with a contribution to the backward Bose enhancement of the gluons $k_2-q_2$ and $k_3-q_3$ in the projectile wave function. The second piece of $I_{\rm 2tr,3}^{(2)}$ is proportional to 
\beq
\label{mu_2tr-3-2-2}
&&
\mu^2\big[ k_3-q_2,q_3-k_3\big] 
\bigg\lgroup  \mu^2\big[ k_1-q_1,q_1-k_2\big]  \mu^2\big[ q_2-k_1,k_2-q_3\big] \nonumber \\
&& \hskip 3.3cm +  \frac{1}{2}  \mu^2\big[ k_1-q_1,k_2-q_3\big]  \mu^2\big[ q_2-k_1,q_1-k_2\big] \bigg\rgroup 
\nonumber\\
&
\propto&
F\big[ |q_2-q_3|R\big] \bigg\lgroup F^2\big[ |k_1-k_2|R\big]\, +\, \frac{1}{2} F^2\big[ |(k_1-q_1)+(k_2-q_2)|R\big]\bigg\rgroup.
\eeq
The first term in this equation is a contribution to the forward Bose enhancement of the gluons $q_2$ and $q_3$ in the target wave function with a forward contribution to HBT of gluons $k_1$ and $k_2$. The second term in Eq.~\eqref{mu_2tr-3-2-2} is a contribution to the forward Bose enhancement of the gluons $q_2$ and $q_3$ in the target wave function with a backward contribution to the Bose enhancement of the gluons $k_1-q_1$ and $k_2-q_2$ in the projectile wave function. Finally, the third piece of $I_{\rm 2tr,3}^{(2)}$ is proportional to 
\beq
\label{mu_2tr-3-3-2}
&&
\mu^2\big[ k_2-q_3,q_2-k_2\big] 
\bigg\lgroup  \mu^2\big[ k_1-q_2,q_3-k_3\big] \mu^2\big[q_1-k_1,k_3-q_1\big] \nonumber \\
&& \hskip 3.3cm +  \frac{1}{2}  \mu^2\big[ k_1-q_2,k_3-q_1\big]  \mu^2\big[ q_1-k_1,q_3-k_3\big] \bigg\rgroup  
\nonumber\\
&
\propto &
F\big[ |q_2-q_3|R\big] \bigg\lgroup F^2\big[ |k_1-k_3|R\big]\, +\, \frac{1}{2}F^2\big[ |(k_1-q_1)+(k_3-q_3)|R\big] \bigg\rgroup.
\eeq
The first term here is a contribution to the forward Bose enhancement of the gluons $q_2$ and $q_3$ in the target wave function together with a contribution to the forward HBT of the gluons $k_1$ and $k_3$. The second term in Eq.~\eqref{mu_2tr-3-3-2} is a contribution to the forward Bose enhancement of the gluons $q_2$ and $q_3$ in the target wave function together with a contribution to backward Bose enhancement to the gluons $k_1-q_1$ and $k_3-q_3$ in the projectile wave function. The symmetry partner of the $I_{\rm 2tr,3}^{(2)}$ that is defined in Eq. \eqref{2-tr-3-2} can be identified easily in the same manner. 
\end{itemize}

\item[(iii)] Finally, we can analyze the terms that are originate from the single-trace contribution. They are four of them:
\begin{itemize}
\item[(a)] The first one, $I_{\rm 1tr,1}^{(2)}$, is defined in Eq.~\eqref{tilde_1tr-1-2} with its symmetry partners given in Eq.~\eqref{1tr-1-2}.
The first term is proportional to 
\beq
\label{mu_1-tr-2-1-2}
&&
\mu^2\big[ k_1-q_1,k_2-q_2\big] 
 \bigg\lgroup \mu^2\big[k_3-q_3,q_1-k_1\big] \mu^2\big[q_2-k_2,q_3-k_3\big] \\
&& \hskip 3.3cm +\mu^2\big[k_3-q_3,q_2-k_2\big] \mu^2\big[q_1-k_1,q_3-k_3\big] \bigg\rgroup
 \nonumber\\
 &
 \propto&
 F\big[ |(k_1-q_1)+(k_2-q_2)|R\big]
 \bigg\lgroup F\big[ |(k_3-q_3)-(k_1-q_1)|R\big] \,  F\big[ |(k_2-q_2)+(k_3-q_3)|R\big]
 \nonumber\\
 &&
 \hspace{4.5cm}
 +\, 
 F\big[ |(k_3-q_3)-(k_2-q_2)|R\big] \, F\big[ |(k_1-q_1)+(k_3-q_3)|R\big]
 \bigg\rgroup.\nonumber
 \eeq
Clearly, the first term in this equation is a contribution to backward Bose enhancement of the gluons $k_1-q_1$ and $k_2-q_2$ together with contribution to forward Bose enhancement of the gluons $k_1-q_1$ and $k_3-q_3$ as well as a contribution to backward Bose enhancement of the gluons $k_2-q_2$ and $k_3-q_3$, all in the projectile wave function. The second term in Eq.~\eqref{mu_1-tr-2-1-2} is a contribution to backward Bose enhancement of the gluons $k_1-q_1$ and $k_2-q_2$ together with a contribution to forward Bose enhancement of the gluons $k_3-q_3$ and $k_2-q_2$ as well as a contribution to backward Bose enhancement of the gluons $k_1-q_1$ and $k_3-q_3$, all in the projectile wave function. The symmetry partners of this term are given in Eq.~\eqref{1tr-1-2} and, again, they can be easily identified by using the same procedure.    

\item[(b)] The second term that originates from the single-trace contribution, $I_{\rm 1tr,2}^{(2)}$, is defined in Eq.~\eqref{tilde_1tr-2-2} with its symmetric partners given in Eq.~\eqref{1tr-2-2}. This term has four pieces and the first piece is proportional to 
\beq
\label{mu_1tr-2-1-2}
&&
\mu^2\big[ k_1-q_2,k_2-q_1\big]  
\bigg\lgroup   \frac{1}{2}\mu^2\big[k_3-q_3,q_1-k_1\big]   \mu^2\big[q_2-k_2,q_3-k_3\big]   \\
&& \hskip 3.3cm+  \frac{1}{2} \mu^2\big[ k_3-q_3,q_2-k_2\big]   \mu^2\big[q_1-k_1,q_3-k_3\big] \bigg\rgroup
\nonumber\\
&
\propto&
F\big[ |(k_1-q_1)+(k_2-q_2)|R\big] 
\bigg\lgroup \frac{1}{2}\, F\big[ |(k_3-q_3)-(k_1-q_1)|R\big] \, F\big[ |(k_2-q_2)+(k_3-q_3)|R\big]
\nonumber\\
&&
 \hspace{4.5cm}
+\, \frac{1}{2}\, F\big[ |(k_3-q_3)-(k_2-q_2)|R\big] \, F\big[ (k_1-q_1)+(k_3-q_3)\big] \bigg\rgroup.\nonumber 
\eeq  
The first term in this equation is a contribution to backward Bose enhancement of the gluons $k_1-q_1$ and $k_2-q_2$ as well as $k_2-q_2$ and $k_3-q_3$ in the projectile wave function together with a forward contribution to Bose enhancement of the gluons $k_1-q_1$ and $k_3-q_3$ in the projectile wave function. The second term in Eq.~\eqref{mu_1tr-2-1-2} is a contribution to backward Bose enhancement of the gluons $k_1-q_1$ and $k_2-q_2$ as well as $k_1-q_1$ and $k_3-q_3$ in the projectile wave function together with a forward contribution to Bose enhancement of the gluons $k_2-q_2$ and $k_3-q_3$ in the projectile wave function. The second piece of $I_{\rm 1tr,2}^{(2)}$ is proportional to 
\beq
\label{mu_2tr-2-2-2}
&&
\mu^2\big[ k_1-q_1,k_3-q_3\big] 
\bigg\lgroup  \mu^2\big[ k_2-q_2,q_2-k_1\big]    \mu^2\big[q_1-k_2,q_3-k_3\big]    \\
&& \hskip 3.3cm+  \frac{1}{2} \mu^2\big[ k_2-q_2,q_3-k_3\big]  \mu^2\big[ q_2-k_1,q_1-k_2\big] \bigg\rgroup
\nonumber\\
&
\propto &
F\big[ |(k_1-q_1)+(k_3-q_3)|R\big] 
\bigg\lgroup F\big[ |k_1-k_2|R\big] \, F\big[ |(k_1-q_1)+(k_3-q_3)|R\big]
\nonumber\\
&&
 \hspace{4.5cm}
+\, \frac{1}{2}\, F\big[ |(k_2-q_2)-(k_3-q_3)|R\big] \, F\big[ |(k_1-q_1)+(k_2-q_2)|R\big] \bigg\rgroup.\nonumber
\eeq
The first term here is a contribution to backward Bose enhancement of the gluons $k_1-q_1$ and $k_3-q_3$ in the projectile wave function together with a contribution to forward HBT of the gluons $k_1$ and $k_2$. The second term in Eq.~\eqref{mu_2tr-2-2-2} is a contribution to backward Bose enhancement of the gluons $k_1-q_1$ and $k_3-q_3$  as well as the gluons $k_1-q_1$ and $k_2-q_2$ in the projectile wave function together with a contribution to forward Bose enhancement of the gluons $k_2-q_2$ and $k_3-q_3$ in the projectile wave function. The third piece of $I_{\rm 1tr,2}^{(2)}$ is proportional to
\beq
\label{mu_2tr-2-3-2}
&&
\mu^2\big[ k_1-q_1,q_1-k_2\big]
\bigg\lgroup   \mu^2\big[ k_2-q_2,k_3-q_3\big]  \mu^2\big[q_2-k_1,q_3-k_3\big] \\
&& \hskip 3.3cm+ \mu^2\big[ k_2-q_2,q_3-k_3\big] \mu^2\big[ k_3-q_3,q_2-k_1\big] \bigg\rgroup
\nonumber\\
&
\propto&
F\big[ |k_1-k_2|R\big] \bigg\lgroup F^2\big[ |(k_2-q_2)+(k_3-q_3)|R\big]\, +\, F^2\big[ |(k_2-q_2)-(k_3-q_3)|R\big]\bigg\rgroup.
\eeq 
Clearly, the first term this equation is a contribution to forward HBT of the gluons $k_1$ and $k_2$ together with a contribution to backward Bose enhancement of the gluons $k_2-q_2$ and $k_3-q_3$ in the projectile wave function, while the second term is a contribution to forward HBT of the gluons $k_1$ and $k_2$ together with a contribution to forward Bose enhancement of the gluons $k_2-q_2$ and $k_3-q_3$ in the projectile wave function. The last piece of the $I_{\rm 1tr,2}^{(2)}$ is proportional to 
\beq
\label{mu_2tr2-4-2}
&&
\mu^2\big[k_1-q_1,q_3-k_3\big] 
\bigg\lgroup  \frac{1}{2} \mu^2\big[ k_2-q_2,k_3-q_3\big] \mu^2\big[q_2-k_1,q_1-k_2\big] \\
&& \hskip 3.3cm+ \mu^2\big[ k_2-q_2,q_2-k_1\big] \mu^2\big[k_3-q_3,q_1-k_2\big] \bigg\rgroup 
\nonumber\\
&
\propto&
F\big[ |(k_1-q_1)-(k_3-q_3)|R\big]
\bigg\lgroup  \frac{1}{2} \, F\big[ |(k_2-q_2)+(k_3-q_3)|R\big] \, F\big[ |(k_1-q_1)+(k_2-q_2)|\big] 
\nonumber\\
&& 
\hspace{4.5cm}
+\, F\big[ |k_1-k_2|R\big]\, F\big[ |(k_1-q_1)-(k_3-q_3)|R\big] \bigg\rgroup.
\eeq
The first term in this equation is a contribution to the forward Bose enhancement of the gluons $k_1-q_1$ and $k_3-q_3$ together with a contribution to the backward Bose enhancement of the gluons $k_2-q_2$ and $k_3-q_3$ as well as the gluons $k_1-q_1$ and $k_2-q_2$ in the projectile wave function. The second term in Eq.~\eqref{mu_2tr2-4-2} is a contribution to forward Bose enhancement of the gluons  $k_1-q_1$ and $k_3-q_3$ in the projectile wave function together with a contribution to forward HBT of the gluons $k_1$ and $k_2$. The identification of the symmetry partners of $I_{\rm 1tr,2}^{(2)}$ can be performed in a straight forward way by adopting the same procedure. 

\item[(c)] The third term that originates from the single-trace contribution, $I_{\rm 1tr, 3}^{(2)}$ is defined in Eq.~\eqref{tilde_1-tr-3-2} and its symmetry partners are given in Eq.~\eqref{1-tr-3-2}. This term has also four pieces and the first one is proportional to 
\beq
\label{mu_1tr_3-1-2}
&&
\bigg\{
 \mu^2\big[ k_1-q_1,k_2+q_1\big]  
\bigg\lgroup \mu^2\big[ k_3+q_2,-q_2-k_1\big] \mu^2\big[-q_3-k_2,q_3-k_3\big]
 \\
&& \hskip 3.5cm+ \frac{1}{2} \mu^2\big[ k_3+q_2,-q_3-k_2\big]  \mu^2\big[-q_2-k_1,q_3-k_3\big] \bigg\rgroup
 \nonumber\\
 &&
 \hspace{0.3cm}
 +
\frac{1}{4} \mu^2\big[ k_1-q_1,q_3-k_3\big] \mu^2\big[ k_2+q_1,-q_2-k_1\big] \mu^2\big[ k_3+q_2,-q_3-k_2\big] \bigg\}
\nonumber\\
 &
 \propto&
 \bigg\{
 F\big[ |k_1+k_2|R\big]
 \bigg\lgroup F\big[ |k_1-k_3|R\big]\, F\big[ |k_2+k_3|R\big]\, +\, \frac{1}{2}\, F^2\big[ |(k_3-q_3)-(k_2-q_2)|R\big] \bigg\rgroup 
 \nonumber\\
 &&
  \hspace{0.3cm}
 +\frac{1}{4}\, F\big[ |(k_1-q_1)-(k_3-q_3)|R\big]\, F\big[ |(k_1-q_1)-(k_2-q_2)|R\big] \, F\big[ |(k_2-q_2)-(k_3-q_3)|R\big]\bigg\}.\nonumber
\eeq
The first term in this equation is a contribution to the backward HBT of the gluons $k_1$ and $k_2$ as well as the gluons $k_2$ and $k_3$ together with a contribution to forward HBT of the gluons $k_1$ and $k_3$. The second term in Eq.~\eqref{mu_1tr_3-1-2} is a contribution to  backward HBT of the gluons $k_1$ and $k_2$ together with a contribution to forward Bose enhancement of the gluons $k_2-q_2$ and $k_3-q_3$ in the projectile wave function. The third term is a contribution to forward Bose enhancement of the all three gluons $k_1-q_1$, $k_2-q_2$ and $k_3-q_3$ in the projectile wave function. The second piece of $I_{\rm 1tr, 3}^{(2)}$ is proportional to 
\beq
\label{mu_1tr3-2-2}
&&
\bigg\lgroup \mu^2\big[ k_1-q_1, k_3-q_3\big]\, \mu^2\big[ k_2+q_1,q_3-k_1\big]\, \mu^2\big[ q_2-k_2,-k_3-q_2\big]
\\
&&
\hspace{0.3cm}
+\, \mu^2\big[ k_1-q_1, q_2-k_2\big] \, \mu^2\big[ k_2+q_1, -k_3-q_2\big]\, \mu^2\big[ k_3-q_3,q_3-k_1\big]
\bigg\rgroup
\nonumber\\
&
\propto& \bigg\lgroup F\big[ |k_2+k_3|R\big]\, F^2\big[ |(k_1-q_1)+(k_3-q_3)|R\big]\, +\, F\big[ |k_1-k_3|R\big]\, F^2\big[ |(k_1-q_1)-(k_2-q_2)|R\big]\bigg\rgroup.\nonumber
\eeq
The first term here is a contribution to backward HBT of the gluons $k_2$ and $k_3$ together with a contribution to backward Bose enhancement of the gluons $k_1-q_1$ and $k_3-q_3$ in the projectile wave function. The second term in this equation is a contribution to forward HBT of the gluons $k_1$ and $k_3$ together with a contribution to forward Bose enhancement of the gluons $k_1-q_1$ and $k_2-q_2$ in the projectile wave function. The third piece of $I_{\rm 1tr, 3}^{(2)}$ is proportional to 
\beq
&&
 \mu^2\big[ k_1-q_3,k_3-q_1\big]\, \mu^2\big[ k_2+q_3,-k_3-q_2\big]\, \mu^2\big[ q_1-k_1,q_2-k_2\big] 
 \nonumber\\
 &
 \propto &
 F\big[ |(k_1-q_1)+(k_3-q_3)|R\big]\, F\big[ |(k_2-q_2)-(k_3-q_3)|R\big]\, F\big[ |(k_1-q_1)+(k_2-q_2)|R\big].
\eeq
This term is a contribution to backward Bose enhancement of the gluons $k_1-q_1$ and $k_3-q_3$ as well as the gluons $k_1-q_1$ and $k_2-q_2$ in the projectile wave function together with a contribution to forward Bose enhancement of the gluons $k_2-q_2$ and $k_3-q_3$ in the projectile wave function. The last piece of $I_{\rm 1tr, 3}^{(2)}$ is proportional to
\beq
&&
\mu^2\big[ k_1+q_2,-q_1-k_2\big]\, \mu^2\big[ k_2-q_2,k_3-q_3\big]\, \mu^2\big[q_3-k_1,q_1-k_3\big]
\nonumber\\
&
\propto&
F\big[ |(k_1-q_1)-(k_2-q_2)|R\big]\, F\big[ |(k_2-q_2)+(k_3-q_3)|R\big] \, F\big[ |(k_1-q_1)+(k_3-q_3)|R\big].
\eeq
This term is a contribution to the backward Bose enhancement of the gluons $k_1-q_1$ and $k_3-q_3$ as well as the gluons $k_2-q_2$ and $k_3-q_3$ in the projectile wave function together with a contribution to the forward Bose enhancement of the gluons $k_1-q_1$ and $k_2-q_2$ in the projectile wave function. The symmetry partners of $I_{\rm 1tr, 3}^{(2)}$ can be identified in a simillar manner.

\item[(d)] The last term that originates from the single-trace contribution, $I_{\rm 1tr,4}^{(2)}$, is defined in Eq.~\eqref{tilde_1-tr-4-2_P} with its symmetry partners given in Eq. \eqref{1-tr-4-2_P}. This term has four pieces and the first one is proportional to 
\beq
\label{mu_1tr412}
&&
\bigg\{ \mu^2\big[ k_1-q_1,q_1-k_2\big] 
\bigg\lgroup  \frac{1}{2} \mu^2\big[ k_2-q_3,k_3-q_2\big]  \mu^2\big[q_2-k_1,q_3-k_3\big]\\
&& \hskip 3.6cm+ \mu^2\big[ k_2-q_3,q_3-k_3\big]  \mu^2\big[ k_3-q_2,q_2-k_1\big] \bigg\rgroup
\nonumber \\
&&
\hspace{0.3cm}
+
\frac{1}{4}  \mu^2\big[ k_1-q_1,q_3-k_3\big]  
  \mu^2\big[ k_2-q_3,k_3-q_2\big]  \mu^2\big[ q_2-k_1,q_1-k_2\big]
\bigg\} 
\nonumber\\
&
\propto&
\bigg\{ F\big[ |k_1-k_2|R\big]
\bigg\lgroup \frac{1}{2} F^2\big[ |(k_2-q_2)+(k_3-q_3)|R\big] \, +\, F\big[ |k_2-k_3|R\big]\,  F\big[ |k_1-k_3|R\big]\bigg\rgroup
\nonumber\\
&&
\hspace{0.3cm}
+
\frac{1}{4} \, F\big[ |(k_1-q_1)-(k_3-q_3)|R\big]\, F\big[ |(k_2-q_2)+(k_3-q_3)|R\big]\, F\big[ |(k_1-q_1)+(k_2-q_2)|R\big] \bigg\}.\nonumber
\eeq
The first term in this equation is a contribution to the forward HBT of the gluons $k_1$ and $k_2$ together with a contribution to the backward Bose enhancement of the gluons $k_2-q_2$ and $k_3-q_3$ in the projectile wave function. The second term in Eq.~\eqref{mu_1tr412} is a contribution to forward HBT of the three gluons $k_1$, $k_2$ and $k_3$. The last term is a contribution the backward Bose enhancement of the gluons $k_1-q_1$ and $k_2-q_2$ as well as the gluons $k_2-q_2$ and $k_3-q_3$ together with a contribution to the forward Bose enhancement of the gluons $k_1-q_1$ and $k_3-q_3$ in the projectile wave function. The second piece of $I_{\rm 1tr,4}^{(2)}$ is proportional to 
\beq
\label{mu_1tr-3-2-2}
&&
\bigg\lgroup \mu^2\big[k_1-q_1,k_2-q_2\big]  \mu^2\big[ k_3-q_3,q_3-k_1\big]  \mu^2\big[ q_1-k_2,q_2-k_3\big]
 \\
&&
\hspace{0.4cm}
+
\mu^2\big[ k_1-q_1,k_3-q_3\big] \mu^2\big[ k_2-q_2,q_2-k_3\big]  \mu^2\big[ q_3-k_1,q_1-k_2\big]\bigg\rgroup
\nonumber\\
&
\propto&
\bigg\lgroup F^2\big[ |(k_1-q_1)+(k_2-q_2)|R\big] \, F\big[ |k_1-k_3|R\big]\, +\, F^2\big[ |(k_1-q_1)+(k_3-q_3)|R\big] \, F\big[ |k_2-k_3|R\big]\bigg\rgroup.\nonumber
\eeq
The first term in this equation is a contribution to the forward HBT of the gluons $k_1$ and $k_3$ together with a contribution to the backward Bose enhancement of the gluons $k_1-q_1$ and $k_2-q_2$ in the projectile wave function. The second term in Eq. \eqref{mu_1tr-3-2-2} is a contribution to the forward HBT of the gluons $k_1$ and $k_2$ together with a contribution to the backward Bose enhancement of the gluons $k_1-q_1$ and $k_3-q_3$ in the projectile wave function. The third piece of $I_{\rm 1tr,4}^{(2)}$ is proportional to 
\beq
&&
\mu^2\big[ k_1-q_2,k_2-q_1\big] \mu^2\big[ k_3-q_3,q_2-k_2\big] \mu^2\big[ q_3-k_1, q_1-k_3\big] 
\nonumber\\
&
\propto&
F\big[ |(k_1-q_1)+(k_2-q_2)|R\big] \, F\big[|(k_3-q_3 )-(k_2-q_2)|R\big] \, F\big[ |(k_3-q_3)+(k_1-q_1)|R\big].
\eeq
This term is a contribution to the backward  Bose enhancement of the gluons $k_1-q_1$ and $k_2-q_2$ as well as the gluons $k_1-q_1$ and $k_3-q_3$ together with a contribution to forward Bose enhancement of the gluons $k_2-q_2$ and $k_3-q_3$ in the projectile wave function. The last piece of $I_{\rm 1tr,4}^{(2)}$ is proportional to
\beq
&&
\mu^2\big[ k_1-q_3,k_3-q_1\big] \mu^2\big[ k_2-q_2,q_1-k_1\big] \mu^2\big[ q_3-k_2, q_2-k_3\big]  
\nonumber\\
&
\propto&
F\big[ |(k_1-q_1)+(k_3-q_3)|R\big] \, F\big[|(k_1-q_1 )-(k_2-q_2)|R\big] \, F\big[ |(k_2-q_2)+(k_3-q_3)|R\big].
\eeq 
This term is a contribution to the backward Bose enhancement of the gluons $k_1-q_1$ and $k_3-q_3$ as well as the gluons $k_2-q_2$ and $k_3-q_3$ together with a contribution to the forward Bose enhancement of the gluons $k_1-q_1$ and $k_2-q_2$. The symmetry partners to $I_{\rm 1tr,4}^{(2)}$ can be identified in a similar way.
\end{itemize}
\end{itemize}
\end{itemize}

\section{Discussion and outlook}
\label{sec:conclu}
To conclude, we have derived the non-eikonal Lipatov vertex that takes into account the finite longitudinal width of the target to all orders. This result was conjectured in~\cite{Altinoluk:2015xuy} after considering the first two corrections to the eikonal limit of the Lipatov vertex coming from  the non-eikonal expansion of the gluon propagagor obtained in~\cite{Altinoluk:2014oxa,Altinoluk:2015gia}. However, here we have presented a different derivation from first principles. Then, we have used the non-eikonal Lipatov vertex to study the single, double and triple inclusive gluon production cross sections in p-A collisions at mid pseudorapidity.  Our results are valid for dilute-dilute collisions since we consider the dilute target limit which, for double and triple inclusive gluon production, corresponds to the original Glasma graph calculation with the exception that we take into account the non-eikonal corrections due to the finite longitudinal thickness of the target.

In the single inclusive gluon production cross section, we have shown that the non-eikonal corrections are encoded in  function ${\cal G}_1^{\rm NE}(k^-,\lambda^+)$ that is defined in Eq.~\eqref{G_1} with $k^-$ being the light cone energy of the produced gluon and $\lambda^+$  the colour correlation length along the longitudinal direction in the target. On the one hand, in the limit of $(k^-\lambda^+)\to 0$, our result reproduces the well known eikonal expression which is often referred to as the $k_t$-factorized formula in the CGC. On the other hand, by expanding our result to second order in $(k^-\lambda^+)$, we recover the result calculated in~\cite{Altinoluk:2015xuy}. Our numerical results show that in the kinematic region where the non-eikonal effects are expected to be sizeable, the relative importance of the non-eikonal corrections can vary from $2$ to $15\%$ with respect to the eikonal result. This shows that, depending on the kinematic region that one is interested in, the non-eikonal effects might very well be sizable. 

We have also used the non-eikonal Lipatov vertex to calculate the double inclusive gluon production cross section for dilute-dilute scattering. Adopting the same strategy that was introduced in~\cite{Altinoluk:2018ogz}, we have identified the terms that contribute to uncorrelated production, those that  are responsible for Bose enhancement of the gluons in the projectile and in the target wave functions, and the terms that contribute to HBT interference effects. Our results agree with the results in~\cite{Altinoluk:2018ogz} up to the $N_c$ counting of the target Bose enhancement and part of the projectile Bose enhancement terms. However, it is known that this difference is a consequence of the fact that some aspects of $N_c$ counting are different in the dilute and dense limits~\cite{Altinoluk:2018ogz,Altinoluk:2014mta,Altinoluk:2014twa}.

Moreover, including the non-eikonal corrections in the double inclusive gluon production cross section has a direct consequence. On top of the function ${\cal G}_1^{\rm NE}(k_1^-; \lambda^+)$ that also exists in the single inclusive case, a new function ${\cal G}_2^{\rm NE}(k_1^-,k_2^-; L^+)$, defined in Eq.~\eqref{G_2}, appears which also encodes non-eikonal effects. The partners of the terms that contain ${\cal G}_2^{\rm NE}(k_1^-,k_2^-; L^+)$, obtained via $(\underline k_2\to -\underline k_2)$,   also appear in the double inclusive gluon production cross section but they are accompanied by ${\cal G}_2^{\rm NE}(k_1^-,-k_2^-; L^+)$. However, in some specific kinematic regions, namely when $k_1^-\sim k_2^-$, ${\cal G}_2^{\rm NE}(k_1^-,k_2^-; L^+)\gg {\cal G}_2^{\rm NE}(k_1^-,-k_2^-; L^+)$  which creates an asymmetry. We would like to emphasize that this asymmetry is absent in the eikonal limit. One can immediately realize that this asymmetry created by the non-eikonal corrections in the double inclusive gluon production indeed mimics the asymmetry between the forward and backward peaks in the ridge observed in the two particle correlations. The consequences of this asymmetry are illustrated in Fig.~\ref{double_inc_1} and in Fig.~\ref{double_inc_2}. This is one of the most striking results of our current study. A dedicated study of two particle correlations and azimuthal harmonics with non-eikonal corrections is left for a forthcoming work~\cite{Azimuthal_Harm_NonEik}. 

Finally, we have also considered the non-eikonal triple inclusive gluon production cross section in the dilute target limit. We have identified all the terms that appear in the final result. Compared to the work performed in~\cite{Altinoluk:2018ogz}, the main difference -- apart from non-eikonal corrections that we have included in our study -- is that we have included all  terms while only the leading $N_c$ ones were considered in~\cite{Altinoluk:2018ogz}. This difference is again due to the fact that $N_c$ counting is different  in the dilute and dense regimes. In our study, we have identified the terms that correlate all three gluons which originate from  three-trace or double-trace contributions, which were absent in~\cite{Altinoluk:2018ogz} since they are suppressed in powers of $N_c$ in the dense target limit and therefore discarded there. Moreover, the non-eikonal effects enter through two new functions ${\cal G}_3(k_1^-,k_2^-,k_3^-; L^+)$ and ${\cal G}_4(k_1^-,k_2^-,k_3^-; L^+)$ that are defined in Eqs.~\eqref{G_3} and \eqref{G_4} respectively, on top of the functions ${\cal G}_1(k^-; \lambda^+)$ and ${\cal G}_2(k_1^-,k_2^-; L^+)$  that already appeared in the double inclusive case. Obviously, in the limit of the vanishing $L^+$ these functions become one and provide the eikonal limit of the triple inclusive gluon production cross section in the dilute target limit.

\section*{Acknowledgements}
We thank Raju Venugopalan for comments on the first version of this manuscript. TA  expresses his gratitude to Instituto Galego de F\'{\i}sica de Altas Enerx\'{\i}as for support and hospitality when part of this work was done. 
PA and NA are supported by  Ministerio de Ciencia e Innovaci\'on of Spain under projects FPA2014-58293-C2-1-P, FPA2017-83814-P and Unidad de Excelencia Mar\'{\i}a de Maetzu under project MDM-2016-0692,  by Xunta de Galicia under project ED431C 2017/07, and by FEDER.
The work of TA is supported by Grant No. 2017/26/M/ST2/01074 of the National Science Centre, Poland.
This work has been performed in the framework of COST Action CA15213 "Theory of hot matter and relativistic heavy-ion collisions" (THOR).

\appendix
\section{Details of the calculation of the triple inclusive gluon cross section beyond the eikonal approximation}
\label{app:triple}

%
As in the case of single and double inclusive gluon production, we first take the dilute target limit which corresponds to the expansion of the Wilson lines in powers of the background field of the target, Eq.~\eqref{expanded_U}. Then the triple inclusive gluon production cross section reads
\beq
&&
\frac{d\sigma}{d^2k_1d\eta_1\, d^2k_2d\eta_2 \, d^2k_3d\eta_3}\bigg|_{\rm dilute}=(4\pi)^3 \, \alpha_s^3
\int_{z_1\bar z_1 z_2\bar z_2 z_3\bar z_3}  e^{ik_1\cdot(z_1-\bar{z}_1)+ik_2\cdot(z_2-\bar{z}_2)+ik_3\cdot(z_3-\bar{z}_3)} 
\int_{x_1x_2x_3y_1y_2y_3}
\nonumber\\
&&
\times\, 
A^i(x_1-z_1) A^i(\bar{z}_1-y_1) A^j(x_2-z_2) A^j(\bar{z}_2-y_2) A^k(x_3-z_3)A^k(\bar{z}_3-y_3) 
 \Big\langle \rho^{a_1}_{x_1}\rho^{a_2}_{x_2}\rho^{a_3}_{x_3}\rho^{b_1}_{y_1}\rho^{b_2}_{y_2}\rho^{b_3}_{y_3}\Big\rangle_P 
 \nonumber\\
 &&
 \times\, 
 g^6 \int dx_1^+ dx_2^+ dx_3^+ dx_4^+ dx_5^+ dx_6^+ 
 \int \frac{d^2q_1}{(2\pi)^2} \frac{d^2q_2}{(2\pi)^2} \frac{d^2q_3}{(2\pi)^2} \frac{d^2q_4}{(2\pi)^2} \frac{d^2q_5}{(2\pi)^2} \frac{d^2q_6}{(2\pi)^2}
\left( T^{c_1} T^{c_2}\right)_{a_1 b_1} \left( T^{c_3} T^{c_4}\right)_{a_2 b_2} \left( T^{c_5} T^{c_6}\right)_{a_3 b_3}
\nonumber\\
&&
\times\, 
\Big\langle 
A^-_{c_1}(x_1^+,q_1)A^-_{c_2}(x_2^+,q_2) A^-_{c_3}(x_3^+,q_3) A^-_{c_4}(x_4^+,q_4) A^-_{c_5}(x_6^+,q_6)
\Big\rangle_T \; 
\Big[ e^{-iq_1\cdot z_1}- e^{-iq_1\cdot x_1} \Big] \Big[e^{iq_2\cdot \bar z_1}- e^{iq_2\cdot y_1}\Big]
\nonumber\\
&&
\times\; 
\Big[ e^{-iq_3\cdot z_2}- e^{-iq_3\cdot x_2} \Big] \Big[e^{iq_4\cdot \bar z_2}- e^{iq_4\cdot y_2}\Big]
\Big[ e^{-iq_5\cdot z_3}- e^{-iq_5\cdot x_3} \Big] \Big[e^{iq_6\cdot \bar z_3}- e^{iq_6\cdot y_3}\Big].
\eeq
In the calculation of the single and double inclusive gluon production cross section, we performed the averaging over the colour charge densities of the projectile first. However, it can also be left for further stages of the calculation for convenience since the expressions for the triple inclusive gluon production are longer. Therefore, we leave it for later and perform the integrals over the transverse coordinates which yields 
\beq
\label{triple_eik}
&&
\frac{d\sigma}{d^2k_1d\eta_1\, d^2k_2d\eta_2 \, d^2k_3d\eta_3}\bigg|_{\rm dilute}=(4\pi)^3 \, \alpha_s^3\, g^6\,  \int dx_1^+ dx_2^+ dx_3^+ dx_4^+ dx_5^+ dx_6^+ 
\int \frac{d^2q_1}{(2\pi)^2} \frac{d^2q_2}{(2\pi)^2} \frac{d^2q_3}{(2\pi)^2} \frac{d^2q_4}{(2\pi)^2} \frac{d^2q_5}{(2\pi)^2} \frac{d^2q_6}{(2\pi)^2}
\nonumber\\
&&
\times\; 
\Big\langle 
A^-_{c_1}(x_1^+,q_1)A^-_{c_2}(x_2^+,q_2) A^-_{c_3}(x_3^+,q_3) A^-_{c_4}(x_4^+,q_4) A^-_{c_5}(x_6^+,q_6)
\Big\rangle_T \; 
\Big\langle 
\rho^{a_1}_{k_1-q_1} \, \rho^{a_2}_{k_2-q_3} \, \rho^{a_3}_{k_3-q_5} \, 
\rho^{b_1}_{q_2-k_1} \, \rho^{b_2}_{q_4-k_2} \, \rho^{b_3}_{q_6-k_3} 
\Big\rangle_P
\nonumber\\
&&
\times\; 
\left( T^{c_1} T^{c_2}\right)_{a_1 b_1} \left( T^{c_3} T^{c_4}\right)_{a_2 b_2} \left( T^{c_5} T^{c_6}\right)_{a_3 b_3}
L^i(k_1,q_1)L^i(k_1,q_2)\, L^j(k_2,q_3)L^j(k_2,q_4)\, L^k(k_3,q_5)L^k(k_3,q_6),
\eeq
where $L^i(k,q)$ is the eikonal Lipatov vertex defined in Eq. \eqref{EikL}. At this point, we can incorporate the non-eikonal effects for the triple inclusive gluon production cross section. As discussed earlier, these effects are taken into account by exchanging each eikonal Lipatov vertex in Eq. \eqref{triple_eik} with the corresponding non-eikonal Lipatov vertex given in Eq. \eqref{NEikL}, and using Eq. \eqref{correlation} for the correlator of two target fields. After exchanging each eikonal Lipatov vertex with the corresponding non-eikonal one, the dilute target limit of the non-eikonal triple inclusive gluon production cross section reads
\beq
\label{triple_non-eik}
&&
\frac{d\sigma}{d^2k_1d\eta_1\, d^2k_2d\eta_2 \, d^2k_3d\eta_3}\bigg|_{\rm dilute}^{\rm NE}=(4\pi)^3 \, \alpha_s^3\, g^6\,  
\int \frac{d^2q_1}{(2\pi)^2} \frac{d^2q_2}{(2\pi)^2} \frac{d^2q_3}{(2\pi)^2} \frac{d^2q_4}{(2\pi)^2} \frac{d^2q_5}{(2\pi)^2} \frac{d^2q_6}{(2\pi)^2}
\int dx_1^+ dx_2^+ dx_3^+ dx_4^+ dx_5^+ dx_6^+ 
\nonumber\\
&&
\times\, 
e^{ik_1^-(x_1^+-x_2^+)+ik_2^-(x_3^+-x_4^+)+ik_3(x_5^+-x_6^+)}
\Big\langle 
A^-_{c_1}(x_1^+,q_1)A^-_{c_2}(x_2^+,q_2) A^-_{c_3}(x_3^+,q_3) A^-_{c_4}(x_4^+,q_4) A^-_{c_5}(x_6^+,q_6)
\Big\rangle_T 
\nonumber\\
&&
\times\; 
\Big\langle 
\rho^{a_1}_{k_1-q_1} \, \rho^{a_2}_{k_2-q_3} \, \rho^{a_3}_{k_3-q_5} \, 
\rho^{b_1}_{q_2-k_1} \, \rho^{b_2}_{q_4-k_2} \, \rho^{b_3}_{q_6-k_3} 
\Big\rangle_P
\; 
\left( T^{c_1} T^{c_2}\right)_{a_1 b_1} \left( T^{c_3} T^{c_4}\right)_{a_2 b_2} \left( T^{c_5} T^{c_6}\right)_{a_3 b_3}
\nonumber\\
&&
\times\; 
L^i(k_1,q_1)L^i(k_1,q_2)\, L^j(k_2,q_3)L^j(k_2,q_4)\, L^k(k_3,q_5)L^k(k_3,q_6).
\eeq
Let us now consider the averaging over the colour fields of the target. As in the case of the double inclusive gluon production, the average over six colour fields of the target can be factorized into all possible Wick contractions: 
\begin{align}\label{wick2}
\Braket{A^-_1 A^-_2 A^-_3 A^-_4 A^-_5 A^-_6}_T&=\Braket{A^-_1 A^-_2}_T \left[ \Braket{A^-_3 A^-_4}_T \Braket{A^-_5 A^-_6}_T+\Braket{A^-_3 A^-_5}_T \Braket{A^-_4 A^-_6}_T+\Braket{A^-_3 A^-_6}_T \Braket{A^-_4 A^-_5}_T  \right] \nonumber \\
&+\Braket{A^-_1 A^-_3}_T \left[ \Braket{A^-_2 A^-_4}_T \Braket{A^-_5 A^-_6}_T+\Braket{A^-_2 A^-_5}_T \Braket{A^-_4 A^-_6}_T+\Braket{A^-_2 A^-_6}_T \Braket{A^-_4 A^-_5}_T  \right] \nonumber \\
&+\Braket{A^-_1 A^-_4}_T \left[ \Braket{A^-_2 A^-_3}_T \Braket{A^-_5 A^-_6}_T+\Braket{A^-_2 A^-_5}_T \Braket{A^-_3 A^-_6}_T+\Braket{A^-_2 A^-_6}_T \Braket{A^-_3 A^-_5}_T  \right] \nonumber \\
&+\Braket{A^-_1 A^-_5}_T \left[ \Braket{A^-_2 A^-_3}_T \Braket{A^-_4 A^-_6}_T+\Braket{A^-_2 A^-_4}_T \Braket{A^-_3 A^-_6}_T+\Braket{A^-_2 A^-_6}_T \Braket{A^-_3 A^-_4}_T  \right] \nonumber \\
&+\Braket{A^-_1 A^-_6}_T \left[ \Braket{A^-_2 A^-_3}_T \Braket{A^-_4 A^-_5}_T+\Braket{A^-_2 A^-_4}_T \Braket{A^-_3 A^-_5}_T+\Braket{A^-_2 A^-_5}_T \Braket{A^-_3 A^-_4}_T  \right],
\end{align}
where we have introduced a shorthand notation for the target fields $A^-_i\equiv A^-_{c_i}(x_i^+,q_i)$ for convenience. The target fields are originating from the expansion of the Wilson line in the amplitude (complex conjugate amplitude) when the subscript $i$ is odd (even). With this shorthand notation, the correlator of two target fields defined in Eq.~\eqref{correlation}, can be written in the most convenient way as
\beq
\label{AA_short}
\Braket{A^-_iA^-_j}_T=n(x_i^+)\frac{1}{2\lambda^+}\Theta\Big(\lambda^+-\big|x_i^+-x_j^+\big|\Big)\, \Delta^{ij},
\eeq
where $\Delta^{ij}$ is defined as 
\beq
\label{delta}
\Delta^{ij}=\delta^{c_ic_j} \, (2\pi)^2 \, \delta^{(2)}\Big[ q_i+(-1)^{i+j}q_j\Big] \, \big| a(q_i)\big|^2.
\eeq
%
Note that Eq.~\eqref{triple_non-eik} can now be integrated over the longitudinal coordinates. After plugging the factorized expression for averaging of the colour fields of the target given in Eq.~\eqref{wick2} into Eq.~\eqref{triple_non-eik}, the longitudinal coordinate dependent part of the dilute target limit of the non-eikonal triple inclusive gluon production cross section can be written as 
\beq
\label{3gluon_long-Int}
&&
\int  dx_1^+ dx_2^+ dx_3^+ dx_4^+ dx_5^+ dx_6^+ \; e^{ik_1^-(x_1^+-x_2^+)+ik_2^-(x_3^+-x_4^+)+ik_3(x_5^+-x_6^+)} \Braket{A^-_1 A^-_2 A^-_3 A^-_4 A^-_5 A^-_6}_T
\nonumber\\
&&
=
{\cal G}_1^{\rm NE}(k_1^-; \lambda^+) \,  {\cal G}_1^{\rm NE}(k_2^-; \lambda^+) \,  {\cal G}_1^{\rm NE}(k_3^-; \lambda^+) 
\bigg\{ \Delta^{12}\Delta^{34}\Delta^{56} 
\nonumber\\
&&
+ \;
\Delta^{12} \Big[ {\cal G}_2^{\rm NE}(k_2^-,-k_3^-; L^+) \Delta^{35}\Delta^{46} + {\cal G}_2^{\rm NE}(k_2^-,k_3^-; L^+)\Delta^{36}\Delta^{45}\Big]
\nonumber\\
&&
+\,  \Delta^{34}\Big[ {\cal G}_2^{\rm NE}(k_1^-,-k_3^-; L^+) \Delta^{15}\Delta^{26}+{\cal G}_2^{\rm NE}(k_1^-,k_3^-; L^+)\Delta^{16}\Delta^{25}\Big]
\nonumber\\
&&
+\;  \Delta^{56}\Big[ {\cal G}_2^{\rm NE}(k_1^-,-k_2^-; L^+) \Delta^{13}\Delta^{24}+{\cal G}_2^{\rm NE}(k_1^-,k_2^-; L^+)\Delta^{14}\Delta^{23}\Big]
\nonumber\\
&&
+\; 
{\cal G}_3^{\rm NE}(k_1^-,k_2^-,k_3^-; L^+) \Big[ \Delta^{13}\Delta^{25}\Delta^{46}+ \Delta^{16}\Delta^{24}\Delta^{35}\Big]
+
{\cal G}_3^{\rm NE}(k_2^-,k_1^-,k_3^-; L^+) \Big[ \Delta^{13}\Delta^{26}\Delta^{45} + \Delta^{15}\Delta^{24}\Delta^{36}\Big]
\nonumber\\
&&
+\; 
{\cal G}_3^{\rm NE}(k_1^-,k_3^-,k_2^-; L^+) \Big[ \Delta^{14}\Delta^{26}\Delta^{35}+\Delta^{15}\Delta^{23}\Delta^{46} \Big]
+
{\cal G}_4^{\rm NE}(k_1^-,k_2^-,k_3^-; L^+) \Big[  \Delta^{14}\Delta^{25}\Delta^{36}+\Delta^{16}\Delta^{23}\Delta^{45} \Big]\; \bigg\},
\eeq 
where the functions ${\cal G}_1^{\rm NE}(k^-_i; \lambda^+)$ and ${\cal G}_2^{\rm NE}(k^-_i,k^-_j; L^+)$ are the functions that account for non-eikonal effects  and they are defined in Eqs.~\eqref{G_1} and \eqref{G_2}, respectively. Moreover, for the triple inclusive gluon production the longitudinal coordinate integral produces two new functions ${\cal G}_3^{\rm NE}(k_i^-,k_j^-,k_k^-; L^+) $ and ${\cal G}_4^{\rm NE}(k_i^-,k_j^-,k_k^-; L^+)$ that also account for the non-eikonal effects and read 
\beq
&&
{\cal G}_3^{\rm NE}(k_1^-,k^-_2,k^-_3; L^+)=2\, \frac{-\sin\big[ (k_1^++k_2^+)L^+\big]+\sin\big[(k^-_1-k^-_3)L^+\big]+\sin\big[ (k_2^-+k^-_3)L^+\big]}{\big[ (k_1^++k_2^+)L^+\big]\, \big[(k^-_1-k^-_3)L^+\big]\, \big[ (k_2^-+k^-_3)L^+\big] }
\label{G_3}
\eeq
and 
\beq
&&
{\cal G}_4^{\rm NE}(k_1^-,k^-_2,k^-_3; L^+)=\frac{ \sin\Big[ \frac{(k^-_1-k_2^-)}{2} L^+\Big] \, \sin\Big[ \frac{(k_1^- -k_3^-)}{2} L^+\Big] \, \sin\Big[ \frac{(k_2^- -k_3^-)}{2}L^+ \Big] }{\Big[ \frac{(k^-_1-k_2^-)}{2} L^+\Big] \, \Big[ \frac{(k_1^- -k_3^-)}{2} L^+\Big]\, \Big[ \frac{(k_2^- -k_3^-)}{2}L^+ \Big]} \ .
\label{G_4}
\eeq
Both functions go to 1 when we consider the shockwave (eikonal) limit $L^+ \to 0$.

We can now substitute Eq.~\eqref{3gluon_long-Int} into the dilute target limit of the non-eikonal triple inclusive gluon production cross section given in Eq.~\eqref{triple_non-eik}. By using the definition of $\Delta^{ij}$ given in Eq.~\eqref{delta} and integrating over the three transverse momenta, we get
\beq
\label{triple_non-eik_integrated}
&&
\hspace{-0.3cm}
\frac{d\sigma}{d^2k_1d\eta_1\, d^2k_2d\eta_2 \, d^2k_3d\eta_3}\bigg|_{\rm dilute}^{\rm NE}=(4\pi)^3 \, \alpha_s^3\, g^6\,  
\int \frac{d^2q_1}{(2\pi)^2} \frac{d^2q_2}{(2\pi)^2} \frac{d^2q_3}{(2\pi)^2}\,   \big| a(q_1)\big|^2\,   \big| a(q_2)\big|^2\, \big| a(q_3)\big|^2
\nonumber\\
&&
\hspace{-0.3cm}
\times\, 
{\cal G}_1^{\rm NE}(k_1^-,\lambda^+) {\cal G}_1^{\rm NE}(k_2^-,\lambda^+) {\cal G}_1^{\rm NE}(k_3^-,\lambda^+)
\nonumber\\
&&
\hspace{-0.3cm}
\times\, 
\bigg\{ C_A^3
\Big\langle\rho^{a}_{k_1-q_1}\rho^{b}_{k_2-q_2}\rho^{c}_{k_3-q_3}\rho^{a}_{q_1-k_1}\rho^{b}_{q_2-k_2}\rho^{c}_{q_3-k_3}\Big\rangle_P 
L^i(k_1,q_1)  L^i(k_1,q_1)  L^j(k_2,q_2)  L^j(k_2,q_2)  L^k(k_3,q_3)  L^k(k_3,q_3)
\nonumber\\
&&
\hspace{-0.3cm}
+\bigg\lgroup\!\!
\Big[ 
{\cal G}_2^{\rm NE}(k_1^-,k_2^-; L^+) \, C_A  (T^aT^b)_{a_1 b_1} (T^bT^a)_{a_2 b_2} \delta_{a_3 b_3} L^i(k_1,q_1)  L^i(k_1,q_2)  L^j(k_2,q_1)  L^j(k_2,q_2)  L^k(k_3,q_3)  L^k(k_3,q_3)
\nonumber\\
&&
\hspace{0.2cm}
\times\, 
\Big\langle\rho^{a_1}_{k_1-q_1}\rho^{a_2}_{k_2-q_2}\rho^{a_3}_{k_3-q_3}\rho^{b_1}_{-k_1+q_2}\rho^{b_2}_{-k_2+q_1}\rho^{b_3}_{-k_3+q_3}\Big\rangle_P
+\left(\underline{k}_2 \rightarrow -\underline{k}_2 \right) \Big] + \left(\underline{k}_1 \leftrightarrow \underline{k}_3 \right)+\left(\underline{k}_2 \leftrightarrow \underline{k}_3 \right) 
\bigg\rgroup
\nonumber\\
&&
\hspace{-0.3cm}
+\!
\bigg\lgroup\!\!
\Big[
{\cal G}_3^{\rm NE}(k_1^-,k_2^-,k_3^-; L^+) (T^aT^b)_{a_1 b_1} (T^aT^c)_{a_2 b_2} (T^bT^c)_{a_3 b_3} L^i(k_1,q_1)  L^i(k_1,q_2)  L^j(k_2,-q_1)  L^j(k_2,q_3)  L^k(k_3,q_2)  L^k(k_3,-q_3)
\nonumber\\
&&
\hspace{0.2cm}
\times\, 
\Big\langle\rho^{a_1}_{k_1-q_1}\rho^{a_2}_{k_2+q_1}\rho^{a_3}_{k_3-q_2}\rho^{b_1}_{-k_1+q_2}\rho^{b_2}_{-k_2+q_3}\rho^{b_3}_{-k_3-q_3}\Big\rangle_P+\left(\underline{k}_3 \rightarrow -\underline{k}_3 \right) \Big] + \left(\underline{k}_1 \leftrightarrow \underline{k}_3 \right)+\left(\underline{k}_2 \leftrightarrow \underline{k}_3 \right)
\bigg\rgroup
\nonumber\\
&&
\hspace{-0.3cm}
+\!
\bigg\lgroup\!\!
\Big[
{\cal G}_4^{\rm NE}(k_1^-,k_2^-,k_3^-; L^+) (T^aT^b)_{a_1 b_1} (T^cT^a)_{a_2 b_2} (T^bT^c)_{a_3 b_3} L^i(k_1,q_1)  L^i(k_1,q_2)  L^j(k_2,q_1)  L^j(k_2,q_3)  L^k(k_3,q_2)  L^k(k_3,q_3) 
\nonumber\\
&&
\hspace{0.2cm}
\times\, 
\Big\langle\rho^{a_1}_{k_1-q_1}\rho^{a_2}_{k_2-q_3}\rho^{a_3}_{k_3-q_2}\rho^{b_1}_{-k_1+q_2}\rho^{b_2}_{-k_2+q_1}\rho^{b_3}_{-k_3+q_3}\Big\rangle_P + \left(\underline{k}_2 \leftrightarrow \underline{k}_3 \right) \Big] 
\bigg\rgroup
\bigg\},
\eeq
where we remind the notation $\underline{k} \equiv (k^-,k)$.
Our next order of business is to perform the averaging over the projectile colour charge densities. As in the previous subsections, we adopt the generalized MV model for the average of two projectile colour charge densities and  write down all possible Wick contractions of their products. Then, the average of six generic projectile colour charge densities can be written  
\beq
\label{proj_contraction}
&&
\hspace{-0.4cm}
\Big\langle \rho^{a_1}_{k_1}\rho^{a_2}_{k_2}\rho^{a_3}_{k_3} \rho^{b_1}_{p_1}\rho^{b_2}_{p_2}\rho^{b_3}_{p_3} \Big\rangle _P=
  \big\langle \rho^{a_1}_{k_1} \rho^{b_1}_{p_1}  \big\rangle \big\langle \rho^{a_2}_{k_2} \rho^{b_2}_{p_2} \big\rangle \big\langle \rho^{a_3}_{k_3} \rho^{b_3}_{p_3} \big\rangle
+\; \big\langle \rho^{a_1}_{k_1} \rho^{b_1}_{p_1}  \big\rangle
 \Big[ \big\langle \rho^{a_2}_{k_2} \rho^{a_3}_{k_3} \big\rangle \big\langle \rho^{b_2}_{p_2} \rho^{b_3}_{p_3} \big\rangle
 + \big\langle \rho^{a_2}_{k_2} \rho^{b_3}_{p_3} \big\rangle \big\langle \rho^{a_3}_{k_3} \rho^{b_2}_{p_2} \big\rangle \Big]
\nonumber\\
&&
\hspace{-0.4cm}
+ \, \big\langle \rho^{a_2}_{k_2} \rho^{b_2}_{p_2}  \big\rangle
 \Big[ \big\langle \rho^{a_1}_{k_1} \rho^{a_3}_{k_3} \big\rangle \big\langle \rho^{b_1}_{p_1} \rho^{b_3}_{p_3} \big\rangle
 + \big\langle \rho^{a_1}_{k_1} \rho^{b_3}_{p_3} \big\rangle \big\langle \rho^{a_3}_{k_3} \rho^{b_1}_{p_1} \big\rangle \Big]
 +  \, \big\langle \rho^{a_3}_{k_3} \rho^{b_3}_{p_3}  \big\rangle
 \Big[ \big\langle \rho^{a_1}_{k_1} \rho^{a_2}_{k_2} \big\rangle \big\langle \rho^{b_1}_{p_1} \rho^{b_2}_{p_2} \big\rangle
 + \big\langle \rho^{a_1}_{k_1} \rho^{b_2}_{p_2} \big\rangle \big\langle \rho^{a_2}_{k_2} \rho^{b_1}_{p_1} \big\rangle \Big]
\nonumber\\
&&
\hspace{-0.4cm}
+  \, \big\langle \rho^{a_1}_{k_1} \rho^{a_2}_{k_2}  \big\rangle
 \Big[ \big\langle \rho^{a_3}_{k_3} \rho^{b_1}_{p_1} \big\rangle \big\langle \rho^{b_2}_{p_2} \rho^{b_3}_{p_3} \big\rangle
 + \big\langle \rho^{a_3}_{k_3} \rho^{b_2}_{p_2} \big\rangle \big\langle \rho^{b_1}_{p_1} \rho^{b_3}_{p_3} \big\rangle \Big]
 +  \, \big\langle \rho^{a_2}_{k_2} \rho^{a_3}_{k_3}  \big\rangle
 \Big[ \big\langle \rho^{a_1}_{k_1} \rho^{b_2}_{p_2} \big\rangle \big\langle \rho^{b_1}_{p_1} \rho^{b_3}_{p_3} \big\rangle
 + \big\langle \rho^{a_1}_{k_1} \rho^{b_3}_{p_3} \big\rangle \big\langle \rho^{b_1}_{p_1} \rho^{b_2}_{p_2} \big\rangle \Big]
\nonumber\\
&&
\hspace{-0.4cm}
+  \, \big\langle \rho^{a_2}_{k_2} \rho^{b_1}_{p_1}  \big\rangle
 \Big[ \big\langle \rho^{a_1}_{k_1} \rho^{a_3}_{k_3} \big\rangle \big\langle \rho^{b_2}_{p_2} \rho^{b_3}_{p_3} \big\rangle
 + \big\langle \rho^{a_1}_{k_1} \rho^{b_3}_{p_3} \big\rangle \big\langle \rho^{a_3}_{k_3} \rho^{b_2}_{p_2} \big\rangle \Big]
 +  \, \big\langle \rho^{a_2}_{k_2} \rho^{b_3}_{p_3}  \big\rangle
 \Big[ \big\langle \rho^{a_1}_{k_1} \rho^{b_2}_{p_2} \big\rangle \big\langle \rho^{a_3}_{k_3} \rho^{b_1}_{p_1} \big\rangle
 + \big\langle \rho^{a_1}_{k_1} \rho^{a_3}_{k_3} \big\rangle \big\langle \rho^{b_1}_{p_1} \rho^{b_2}_{p_2} \big\rangle \Big],
\eeq
where the two projectile colour charge correlator is given by Eq.~\eqref{proj_corr}.  One can use Eq.~\eqref{proj_contraction} in order to perform the projectile colour charge density averaging in Eq.~\eqref{triple_non-eik_integrated}. The resulting expression consists of three distinct parts: a term with a single trace, a term with double trace and a term with three traces of the colour generators (these terms are the analogue of three-dipole, dipole-quadrupole and sextuple contributions in~\cite{Altinoluk:2018ogz} for the dilute-dense set up). Therefore, we write the dilute target limit of the non-eikonal triple inclsuive gluon production cross section as sum of those three contributions:

\beq
\label{gen_3_tr}
\frac{d\sigma}{d^2k_1d\eta_1\, d^2k_2d\eta_2 \, d^2k_3d\eta_3}\bigg|_{\rm dilute}^{\rm NE}&=&
\frac{d\sigma^{(\rm 3tr)}}{d^2k_1d\eta_1\, d^2k_2d\eta_2 \, d^2k_3d\eta_3}\bigg|_{\rm dilute}^{\rm NE}+
\frac{d\sigma^{(\rm 2tr)}}{d^2k_1d\eta_1\, d^2k_2d\eta_2 \, d^2k_3d\eta_3}\bigg|_{\rm dilute}^{\rm NE}
\nonumber\\
&
+
&
\frac{d\sigma^{(\rm 1tr)}}{d^2k_1d\eta_1\, d^2k_2d\eta_2 \, d^2k_3d\eta_3}\bigg|_{\rm dilute}^{\rm NE}\ .
\eeq
Let us now write down the explicit expressions for each of these three contributions starting from the the three-trace one:  
\begin{align}
&\frac{d \sigma^{(\rm 3tr)}}{d^2 k_1 d \eta_1 d^2 k_2 d \eta_2 d^2 k_3 d \eta_3} \bigg\rvert_{\rm dilute}^{\rm NE}= (4\pi)^3\, \alpha_s^3\, g^6\, C_A^3 \,(N_c^2-1)^3  
\int \frac{d^2q_1}{(2\pi)^2} \frac{d^2q_2}{(2\pi)^2} \frac{d^2q_3}{(2\pi)^2}
 \big| a(q_1)\big|^2 \,  \big|a(q_2)\big|^2\,   \big|a(q_3)\big|^2 
 \nonumber\\
 &
 \times\, 
 {\cal G}_1^{\rm NE}(k_1^-; \lambda^+)\,  {\cal G}_1^{\rm NE}(k_2^-; \lambda^+) \,  {\cal G}_1^{\rm NE}(k_3^-; \lambda^+)
\Bigg\{  I_{\rm 3tr }^{(0)}+\frac{1}{(N_c^2-1)} I_{\rm 3tr}^{(1)} +\frac{1}{(N_c^2-1)^2} \left[I_{\rm 3tr,1}^{(2)}+I_{\rm 3tr,2}^{(2)}\right] \Bigg\},
\end{align}
where 
\begin{align}
\label{eq:ref0}
I_{\rm 3tr}^{(0)}&=\mu^2\big[ k_1-q_1,q_1-k_1\big] \, \mu^2\big[k_2-q_2,q_2-k_2\big] \mu^2\big[k_3-q_3,q_3-k_3\big] 
\nonumber \\
&\times 
L^i(k_1,q_1)  L^i(k_1,q_1)  L^j(k_2,q_2)  L^j(k_2,q_2)  L^k(k_3,q_3)  L^k(k_3,q_3).
\end{align}
For ${\cal O}\Big( 1/(N_c^2-1)\Big)$ terms, we have introduced the following compact notation 
\beq
\label{3-tr-1-1}
I^{(1)}_{\rm 3tr}= \Big[ {\tilde I^{(1)}_{\rm 3tr}}+\big(\underline k_2\to -\underline k_2\big)\Big]+  \big(\underline{k}_1 \leftrightarrow \underline{k}_3 \big)+\big(\underline{k}_2 \leftrightarrow \underline{k}_3 \big)
\eeq 
with 
\begin{align}
\label{tilde_3-tr-1-1}
{\tilde I^{(1)}_{\rm 3tr}} &= {\cal G}_2^{\rm NE}(k_1^-, k_2^-; L^+) \,  \mu^2\big[ k_1-q_1,q_2-k_1\big]\,  \mu^2\big[k_2-q_2,q_1-k_2\big] \,  \mu^2\big[k_3-q_3,q_3-k_3\big] 
\nonumber\\
&\times\; 
 L^i(k_1,q_1)  L^i(k_1,q_2) \,  L^j(k_2,q_1)  L^j(k_2,q_2) \,  L^k(k_3,q_3)  L^k(k_3,q_3) .
\end{align}
A similar compact notation has been adopted for the ${\cal O}\Big( 1/(N_c^2-1)^2\Big)$ terms in Eq.~\eqref{gen_3_tr}: 
\beq
\label{3-tr-1-2}
I_{\rm 3tr,1}^{(2)}= \Big[ {\tilde I^{(2)}_{\rm 3tr,1}} +\big(\underline k_3\to -\underline k_3\big)\Big] + \big(\underline{k}_1 \leftrightarrow \underline{k}_3 \big)+\big(\underline{k}_2 \leftrightarrow \underline{k}_3 \big)
\eeq
with
\begin{align}
\label{tilde_3-tr-1-2}
{\tilde I_{\rm 3tr,1}^{(2)}}&= 
{\cal G}_3^{\rm NE}(k_1^-,k_2^-,k_3^-; L^+)\, 
 \mu^2\big[k_1-q_1,q_2-k_1\big] \, \mu^2\big[ k_2+q_1,q_3-k_2\big]\,  \mu^2\big[k_3-q_2,-q_3-k_3\big]   
\nonumber \\
&\times 
L^i(k_1,q_1)  L^i(k_1,q_2) \,  L^j(k_2,-q_1)  L^j(k_2,q_3) \,  L^k(k_3,q_2)  L^k(k_3,-q_3),
\end{align}
and
\beq
\label{3-tr-2-2}
I_{\rm 3tr,2}^{(2)}= {\tilde I_{\rm 3tr,2}^{(2)}}+ \big( \underline{k}_2 \leftrightarrow \underline{k}_3\big)
\eeq
with
\begin{align}
\label{tilde_3-tr-2-2}
{\tilde I_{\rm 3tr,2}^{(2)}}&=  {\cal G}_4^{\rm NE}(k_1^-,k_2^-,k_3^-; L^+)\, 
 \mu^2\big[k_1-q_1,q_2-k_1\big] \, \mu^2\big[ k_2-q_3,q_1-k_2\big]\,  \mu^2\big[k_3-q_2,q_3-k_3\big]   \nonumber \\
&\times  L^i(k_1,q_1)  L^i(k_1,q_2) \,  L^j(k_2,q_1)  L^j(k_2,q_3)   \, L^k(k_3,q_2)  L^k(k_3,q_3).
\end{align}

The double-trace contribution to the dilute target limit of the non-eikonal triple inclusive gluon production cross section can be organized in a similar way:  
\begin{align}
&\frac{d \sigma^{(\rm 2tr)}}{d^2 k_1 d \eta_1 d^2 k_2 d \eta_2 d^2 k_3 d \eta_3} \bigg|_{\rm dilute}^{\rm NE}=(4\pi)^3\, \alpha_s^3  \, g^6 \, C_A^3\,  (N_c^2-1)^2  
\int \frac{d^2q_1}{(2\pi)^2} \frac{d^2q_2}{(2\pi)^2}  \frac{d^2q_3}{(2\pi)^2} \, \big|a(q_1)\big|^2 \,  \big|a(q_2)\big|^2\,   \big|a(q_3)\big|^2 
\nonumber \\
&\times\,  
{\cal G}_1^{\rm NE}(k_1^-, \lambda^+) \,  {\cal G}_1^{\rm NE}(k_2^-; \lambda^+) \,  {\cal G}_1^{\rm NE}(k_3^-; \lambda^+) 
\Bigg\{ \Big[ I_{\rm 2tr,1}^{(1)}+I_{\rm 2tr,2}^{(1)}\Big] + \frac{1}{(N_c^2-1)} \left[ I_{\rm 2tr,1}^{(2)}+I_{\rm 2tr,2}^{(2)}+I_{\rm 2tr,3}^{(2)} \right] \Bigg\}.
\end{align}
Similar compact notations can be adopted for each term in the double-trace contribution. Let us start with the ${\cal O}(1)$ terms: 
\beq
\label{2-tr-1-1}
I_{\rm 2tr,1}^{(1)}= \Big[ {\tilde I_{\rm 2tr,1}^{(1)} } + \big(\underline k_2\to -\underline k_2\big) \Big] + \big(\underline{k}_1 \leftrightarrow \underline{k}_2 \big)+\big(\underline{k}_1 \leftrightarrow \underline{k}_3 \big),
\eeq
with 
\begin{align}
\label{tilde_2-tr-1-1}
{\tilde I_{\rm 2tr,1}^{(1)} } &= \mu^2\big[ k_1-q_1,q_1-k_1\big]\,  \mu^2\big[ k_2-q_2,q_3-k_3\big] \,  \mu^2\big[k_3-q_3,q_2-k_2\big]
\nonumber\\
&\times\, 
 L^i(k_1,q_1)  L^i(k_1,q_1) \,  L^j(k_2,q_2)  L^j(k_2,q_2) \,   L^k(k_3,q_3)  L^k(k_3,q_3) ,
\end{align}
and
\beq
\label{2-tr-2-1}
I_{\rm 2tr,2}^{(1)}=\Big[ {\tilde I_{\rm 2tr,2}^{(1)}}+\big( \underline k_2\to-\underline k_2\big)\Big] + \big(\underline{k}_1 \leftrightarrow \underline{k}_3 \big)+\big(\underline{k}_2 \leftrightarrow \underline{k}_3 \big)
\eeq
with 
\begin{align}
\label{tilde_2-tr-2-1}
{\tilde I_{\rm 2tr,2}^{(1)}}&=
{\cal G}_2^{\rm NE}(k_1^-,k_2^-; L^+) \, 
\mu^2\big[ k_3-q_3,q_3-k_3\big] \, L^i(k_1,q_1)  L^i(k_1,q_2) \,  L^j(k_2,q_1)  L^j(k_2,q_2) \,  L^k(k_3,q_3)  L^k(k_3,q_3)
\nonumber\\
&\times\, 
\bigg\{ 
\mu^2\big[ k_1-q_1,q_1-k_2\big] \,  \mu^2\big[ k_2-q_2,q_2-k_1\big] +\frac{1}{2} \mu^2\big[k_1-q_1,k_2-q_2\big]\,  \mu^2\big[ q_2-k_1,q_1-k_2\big] \bigg\}.
\end{align}
${\cal O}\Big(1/(N_c^2-1)\Big)$ terms in the double-trace contribution can be written in a similar manner. The first term reads 
\beq
\label{2-tr-1-2}
I_{\rm 2tr,1}^{(2)}=\Big[ {\tilde  I_{\rm 2tr,1}^{(2)} } + \big( \underline k_2\to -\underline k_2\big) \Big] + \big(\underline{k}_1 \leftrightarrow \underline{k}_3 \big)+\big(\underline{k}_2 \leftrightarrow \underline{k}_3 \big),
\eeq
with 
\begin{align}
\label{tilde_2-tr-1-2}
{\tilde I_{\rm 2tr,1}^{(2)}}&=
{\cal G}_2^{\rm NE}(k_1^-,k_2^-; L^+) \, \mu^2\big[ k_1-q_1,q_2-k_1\big]\, L^i(k_1,q_1)  L^i(k_1,q_2) \,  L^j(k_2,q_1)  L^j(k_2,q_2) \,  L^k(k_3,q_3)  L^k(k_3,q_3)
\nonumber\\
&\times\, 
\Big\{ 
\mu^2\big[ k_2-q_2,q_3-k_3\big] \, \mu^2\big[ k_3-q_2,q_1-k_2\big]  + \mu^2\big[ k_2-q_2,k_3-q_3\big]\, \mu^2\big[ q_1-k_2,q_3-k_3\big] \Big\}.
\end{align}
The second term can be written as
\beq
\label{2-tr-2-2}
I_{\rm 2tr,2}^{(2)}=\Big[ {\tilde  I_{\rm 2tr,2}^{(2)} } + \big( \underline k_3\to -\underline k_3\big) \Big] + \big(\underline{k}_1 \leftrightarrow \underline{k}_3 \big)+\big(\underline{k}_2 \leftrightarrow \underline{k}_3 \big),
\eeq
with 
%

\begin{align}
\label{tilde_2-tr-2-2}
&
{\tilde I_{\rm 2tr,2}^{(2)}}={\cal G}_3^{\rm NE}(k_1^-,k_2^-,k_3^-; L^+)\,  L^i(k_1,q_1)  L^i(k_1,-q_2)\,   L^j(k_2,-q_1)  L^j(k_2,-q_3) \, L^k(k_3,-q_2)  L^k(k_3,q_3) \\
&\times 
 \mu^2\big[ k_1-q_1,-q_2-k_1\big]  
\bigg\lgroup  \frac{1}{2} \mu^2\big[ k_2+q_1,q_3-k_3\big]  \mu^2\big [k_3+q_2,-q_3-k_2\big]  +   \mu^2\big[ k_2+q_1,k_3+q_2\big]  \mu^2\big[ -q_3-k_2,q_3-k_3\big] \bigg\rgroup 
\nonumber\\
&
+
{\cal G}_3^{\rm NE}(k_1^-,k_2^-,k_3^-; L^+)\,  L^i(k_1,q_1)  L^i(k_1,q_3)\,   L^j(k_2,-q_1)  L^j(k_2,q_2) \, L^k(k_3,q_3)  L^k(k_3,-q_2)
\nonumber\\
&
\times
\mu^2\big[ k_2+q_1,q_2-k_2\big] 
\bigg\lgroup \mu^2\big[ k_1-q_1,-k_3-q_2\big]  \mu^2\big[ q_3-k_1,k_3-q_3\big]    +   \frac{1}{2} \mu^2\big[ k_1-q_1,k_3-q_3\big]   \mu^2\big[ q_3-k_1,-k_3-q_2\big] \bigg\rgroup
\nonumber \\
&
+
{\cal G}_3^{\rm NE}(k_1^-,k_2^-,k_3^-; L^+)\,  L^i(k_1,q_1)  L^i(k_1,-q_2)\,   L^j(k_2,-q_1)  L^j(k_2,q_3) \, L^k(k_3,-q_2)  L^k(k_3,-q_3) 
\nonumber\\
&
\times
\mu^2\big[ k_3+q_2,-k_3-q_3\big] 
\bigg\lgroup  \frac{1}{2} \mu^2\big[ k_1-q_1,q_3-k_2\big]  \mu^2\big[ -q_2-k_1,k_2+q_1\big]   +  \mu^2\big[ k_1-q_1,k_2+q_1\big]  \mu^2\big[- q_2-k_1,q_3-k_2\big] \bigg\rgroup.
\nonumber 
\end{align}


Finally, the last term can be written as
\beq
\label{2-tr-3-2}
I_{\rm 2tr,3}^{(2)}= {\tilde I_{\rm 2tr,3}^{(2)} }+\big( \underline k_2 \leftrightarrow \underline k_3\big)
\eeq
with
\begin{align}
\label{tilde_2-tr-3-2}
&
{\tilde I_{\rm 2tr,3}^{(2)}}= 
{\cal G}_4^{\rm NE}(k_1^-,k_2^-,k_3^-; L^+) \,  L^i(k_1,q_1)  L^i(k_1,q_2) \,  L^j(k_2,q_1)  L^j(k_2,q_3) \,  L^k(k_3,q_2)  L^k(k_3,q_3) 
\\
&\times 
\bigg\{  
\mu^2\big[ k_1-q_1,q_2-k_1\big]  
\bigg\lgroup  \mu^2\big[ k_2-q_3,q_3-k_3\big]  \mu^2\big[ k_3-q_2,q_1-k_2\big] +   \frac{1}{2}  \mu^2\big[ k_2-q_3,k_3-q_2\big]  \mu^2\big[ q_1-k_2,q_3-k_3\big]  \bigg\rgroup
\nonumber \\
&
\hspace{0.26cm}
+
\mu^2\big[ k_3-q_2,q_3-k_3\big] 
\bigg\lgroup  \mu^2\big[ k_1-q_1,q_1-k_2\big]  \mu^2\big[ q_2-k_1,k_2-q_3\big] +  \frac{1}{2}  \mu^2\big[ k_1-q_1,k_2-q_3\big]  \mu^2\big[ q_2-k_1,q_1-k_2\big] \bigg\rgroup \bigg\} \nonumber \\
&
\hspace{0.26cm}
+ {\cal G}_4^{\rm NE}(k_1^-,k_2^-,k_3^-; L^+) \,  L^i(k_1,q_1)  L^i(k_1,q_2) \,  L^j(k_2,q_2)  L^j(k_2,q_3) \,  L^k(k_3,q_1)  L^k(k_3,q_3) 
\nonumber
\\
&
\hspace{0.26cm}
\times 
\mu^2\big[ k_2-q_3,q_2-k_2\big] 
\bigg\lgroup  \mu^2\big[ k_1-q_2,q_3-k_3\big] \mu^2\big[q_1-k_1,k_3-q_1\big] +  \frac{1}{2}  \mu^2\big[ k_1-q_2,k_3-q_1\big]  \mu^2\big[ q_1-k_1,q_3-k_3\big] \bigg\rgroup  .
\nonumber
\end{align}

%
The last contribution to the dilute target limit of the non-eikonal triple inclusive gluon production cross section that we need to consider is the single-trace contribution which can be organized as follows: 
\begin{align}
&
\frac{d \sigma^{(\rm 1tr)}}{d^2 k_1 d \eta_1 d^2 k_2 d \eta_2 d^2 k_3 d \eta_3} \bigg|_{\rm dilute}^{\rm NE}= (4\pi)^3 \, \alpha_s^3 \, g^6 \, C_A^3 \, (N_c^2-1)  
\int \frac{d^2q_1}{(2\pi)^2} \frac{d^2q_2}{(2\pi)^2}  \frac{d^2q_3}{(2\pi)^3}   \big|a(q_1)\big|^2  \big|a(q_2)\big|^2  \big|a(q_3)\big|^2 
\nonumber\\
&\times\, 
{\cal G}_1^{\rm NE}(k_1^-; \lambda^+)\,  {\cal G}_1^{\rm NE}(k_2^-; \lambda^+)\,  {\cal G}_1^{\rm NE}(k_3^-; L^+)  \, 
 \Big[ I_{\rm1tr,1}^{(2)}+I_{\rm1tr,2}^{(2)}+I_{\rm1tr,3}^{(2)}+I_{\rm1tr,4}^{(2)}\Big] .
\end{align}
The first term in the single-trace contribution can be written as 
\beq
\label{1tr-1-2}
I_{\rm 1tr,1}^{(2)}=\Big[ {\tilde  I_{\rm 1tr,1} }^{(2)} + \big( \underline k_2\to -\underline k_2\big) \Big] + \big( \underline k_2\leftrightarrow \underline k_3\big)
\eeq
with
\begin{align}
\label{tilde_1tr-1-2}
&{\tilde  I_{\rm 1tr,1} }^{(2)}=L^i(k_1,q_1)  L^i(k_1,q_1)  L^j(k_2,q_2)  L^j(k_2,q_2)  L^k(k_3,q_3)  L^k(k_3,q_3)  \\
&
\hspace{0.26cm}
\times
\mu^2\big[ k_1-q_1,k_2-q_2\big] 
 \bigg\lgroup \mu^2\big[k_3-q_3,q_1-k_1\big] \mu^2\big[q_2-k_2,q_3-k_3\big]+\mu^2\big[k_3-q_3,q_2-k_2\big] \mu^2\big[q_1-k_1+q_1,q_3-k_3\big] \bigg\rgroup.
 \nonumber
\end{align}
In a similar manner, the second term in the single-trace contribution can be written as 
\beq
\label{1tr-2-2}
I_{\rm 1tr,2}^{(2)}=\Big[ {\tilde  I_{\rm 1tr,2} }^{(2)} + \big( \underline k_2\to -\underline k_2\big) \Big] + \big( \underline k_1\leftrightarrow \underline k_3\big) + \big( \underline k_2\leftrightarrow \underline k_3\big)
\eeq
with 
\begin{align}
\label{tilde_1tr-2-2}
&{\tilde I_{\rm 1tr,2}}^{(2)}= 
{\cal G}_2^{\rm NE}(k_1^-,k_2^-; L^+) \,  L^i(k_1,q_1)  L^i(k_1,q_2) \,  L^j(k_2,q_1)  L^j(k_2,q_2) \,  L^k(k_3,q_3)  L^k(k_3,q_3) 
\\
&\times  
\bigg\{ 
\mu^2\big[ k_1-q_2,k_2-q_1\big]  
\bigg\lgroup   \frac{1}{2}\mu^2\big[k_3-q_3,q_1-k_1\big]   \mu^2\big[q_2-k_2,q_3-k_3\big]  +  \frac{1}{2} \mu^2\big[ k_3-q_3,q_2-k_2\big]   \mu^2\big[q_1-k_1,q_3-k_3\big] \bigg\rgroup
\nonumber \\
&
\hspace{0.26cm}
+
\mu^2\big[ k_1-q_1,k_3-q_3\big] 
\bigg\lgroup  \mu^2\big[ k_2-q_2,q_2-k_1\big]    \mu^2\big[q_1-k_2,q_3-k_3\big]   +  \frac{1}{2} \mu^2\big[ k_2-q_2,q_3-k_3\big]  \mu^2\big[ q_2-k_1,q_1-k_2\big] \bigg\rgroup
\nonumber \\
&
\hspace{0.26cm}
+
\mu^2\big[ k_1-q_1,q_1-k_2\big]
\bigg\lgroup   \mu^2\big[ k_2-q_2,k_3-q_3\big]  \mu^2\big[q_2-k_1,q_3-k_3\big] + \mu^2\big[ k_2-q_2,q_3-k_3\big] \mu^2\big[ k_3-q_3,q_2-k_1\big] \bigg\rgroup
\nonumber \\
&
\hspace{0.26cm}
+ 
\mu^2\big[k_1-q_1,q_3-k_3\big] 
\bigg\lgroup  \frac{1}{2} \mu^2\big[ k_2-q_2,k_3-q_3\big] \mu^2\big[q_2-k_1,q_1-k_2\big] + \mu^2\big[ k_2-q_2,q_2-k_1\big] \mu^2\big[k_3-q_3,q_1-k_2\big] \bigg\rgroup \bigg\}.
\nonumber
\end{align}
%
The third term in the single-trace contribution reads
\beq
\label{1-tr-3-2}
I_{\rm 1tr,3}^{(2)}=\Big[ { \tilde  I_{\rm 1tr,3}}^{(2)}+\big( \underline k_3\to -\underline k_3\big) \Big]+ \big( \underline k_1\leftrightarrow \underline k_3\big) +\big( \underline k_2\leftrightarrow \underline k_3\big)
\eeq
with
\begin{align}
\label{tilde_1-tr-3-2}
&{\tilde I_{\rm 1tr,3}}^{(2)}= 
{\cal G}_3(k_1^-,k_2^-,k_3^-; L^+)\,  L^i(k_1,q_1)  L^i(k_1,-q_2) \,  L^j(k_2,-q_1)  L^j(k_2,-q_3)\,   L^k(k_3,-q_2)  L^k(k_3,q_3) 
\\
&
\hspace{0.26cm}
\times
\bigg\{ \mu^2\big[ k_1-q_1,k_2+q_1\big]  
\bigg\lgroup \mu^2\big[ k_3+q_2,-q_2-k_1\big] \mu^2\big[-q_3-k_2,q_3-k_3\big]
\nonumber\\
&
\hspace{5cm}
 + \frac{1}{2} \mu^2\big[ k_3+q_2,-q_3-k_2\big]  \mu^2\big[-q_2-k_1,q_3-k_3\big] \bigg\rgroup
\nonumber \\
&
\hspace{0.56cm}
+
\frac{1}{4} \mu^2\big[ k_1-q_1,q_3-k_3\big] \mu^2\big[ k_2+q_1,-q_2-k_1\big] \mu^2\big[ k_3+q_2,-q_3-k_2\big] \bigg\}
\nonumber \\
&
\hspace{0.26cm}
+
{\cal G}_3(k_1^-,k_2^-,k_3^-; L^+)\,  L^i(k_1,q_1)  L^i(k_1,q_3) \,  L^j(k_2,-q_1)  L^j(k_2,q_2)\,   L^k(k_3,q_3)  L^k(k_3,-q_2) 
\nonumber \\
&
\hspace{0.5cm}
\times
\frac{1}{2}
\bigg\{ \mu^2\big[ k_1-q_1, k_3-q_3\big]\, \mu^2\big[ k_2+q_1,q_3-k_1\big]\, \mu^2\big[ q_2-k_2,-k_3-q_2\big]
\nonumber\\
&
\hspace{1.5cm}
+\, \mu^2\big[ k_1-q_1, q_2-k_2\big] \, \mu^2\big[ k_2+q_1, -k_3-q_2\big]\, \mu^2\big[ k_3-q_3,q_3-k_1\big]
\bigg\}
\nonumber\\
&
\hspace{0.26cm}
+
{\cal G}_3(k_1^-,k_2^-,k_3^-; L^+)\, L^i(k_1,q_3)  L^i(k_1,q_1) \,  L^j(k_2,-q_3)  L^j(k_2,q_2)\,   L^k(k_3,q_1)  L^k(k_3,-q_2) 
\nonumber\\
&
\hspace{0.5cm}
\times\, 
\frac{1}{4}\, \mu^2\big[ k_1-q_3,k_3-q_1\big]\, \mu^2\big[ k_2+q_3,-k_3-q_2\big]\, \mu^2\big[ q_1-k_1,q_2-k_2\big]
\nonumber\\
&
\hspace{0.26cm}
+
{\cal G}_3(k_1^-,k_2^-,k_3^-; L^+)\, L^i(k_1,-q_2)  L^i(k_1,q_3) \,  L^j(k_2,q_2)  L^j(k_2,-q_1)\,   L^k(k_3,q_3)  L^k(k_3,q_1) 
\nonumber\\
&
\hspace{0.5cm}
\times\, 
\frac{1}{4}\, \mu^2\big[ k_1+q_2,-q_1-k_2\big]\, \mu^2\big[ k_2-q_2,k_3-q_3\big]\, \mu^2\big[q_3-k_1,q_1-k_3\big].
\end{align}
%
%
Finally, the last term in the single-trace contribution can be written as 
\beq
\label{1-tr-4-2_P}
I_{\rm 1tr,4}^{(2)}= {\tilde I_{\rm 1tr,4} }^{(2)}+\big( \underline k_2 \leftrightarrow \underline k_3\big)
\eeq
with
\begin{align}
\label{tilde_1-tr-4-2_P}
&{\tilde I_{\rm 1tr,4}}^{(2)}= {\cal G}_4^{\rm NE}(k_1^-,k_2^-,k_3^-; L^+)\,  L^i(k_1,q_1)  L^i(k_1,q_2) \,  L^j(k_2,q_1)  L^j(k_2,q_3) \,  L^k(k_3,q_2)  L^k(k_3,q_3) \\
&
\hspace{0.26cm}
\times
\bigg\{ \mu^2\big[ k_1-q_1,q_1-k_2\big] 
\bigg\lgroup  \frac{1}{2} \mu^2\big[ k_2-q_3,k_3-q_2\big]  \mu^2\big[q_2-k_1,q_3-k_3\big] + \mu^2\big[ k_2-q_3,q_3-k_3\big]  \mu^2\big[ k_3-q_2,q_2-k_1\big] \bigg\rgroup
\nonumber \\
&
\hspace{0.8cm}
+
\frac{1}{4}  \mu^2\big[ k_1-q_1,q_3-k_3\big]  
  \mu^2\big[ k_2-q_3,k_3-q_2\big]  \mu^2\big[ q_2-k_1,q_1-k_2\big]
\bigg\} 
\nonumber\\
&
\hspace{0.26cm}
+
{\cal G}_4^{\rm NE}(k_1^-,k_2^-,k_3^-; L^+)\,  L^i(k_1,q_1)  L^i(k_1,q_3) \,  L^j(k_2,q_1)  L^j(k_2,q_2) \,  L^k(k_3,q_2)  L^k(k_3,q_3)  
\nonumber\\
&
\hspace{0.8cm}
\times \, \frac{1}{2}
\bigg\{ 
\mu^2\big[k_1-q_1,k_2-q_2\big]  \mu^2\big[ k_3-q_3,q_3-k_1\big]  \mu^2\big[ q_1-k_2,q_2-k_3\big]
\nonumber \\
&
\hspace{1.4cm}
+
\mu^2\big[ k_1-q_1,k_3-q_3\big] \mu^2\big[ k_2-q_2,q_2-k_3\big]  \mu^2\big[ q_3-k_1,q_1-k_2\big]\bigg\}
\nonumber \\
&
\hspace{0.26cm}
+
{\cal G}_4^{\rm NE}(k_1^-,k_2^-,k_3^-; L^+)\,  L^i(k_1,q_2)  L^i(k_1,q_3) \,  L^j(k_2,q_1)  L^j(k_2,q_2) \,  L^k(k_3,q_1)  L^k(k_3,q_3)
\nonumber\\
&
\hspace{0.8cm}
\times\, \frac{1}{4}\, 
\mu^2\big[ k_1-q_2,k_2-q_1\big] \mu^2\big[ k_3-q_3,q_2-k_2\big] \mu^2\big[ q_3-k_1, q_1-k_3\big]  
\nonumber\\
&
\hspace{0.26cm}
+
{\cal G}_4^{\rm NE}(k_1^-,k_2^-,k_3^-; L^+)\,  L^i(k_1,q_1)  L^i(k_1,q_3) \,  L^j(k_2,q_2)  L^j(k_2,q_3) \,  L^k(k_3,q_2)  L^k(k_3,q_1)
\nonumber\\
&
\hspace{0.8cm}
\times\, \frac{1}{4}\, 
\mu^2\big[ k_1-q_3,k_3-q_1\big] \mu^2\big[ k_2-q_2,q_1-k_1\big] \mu^2\big[ q_3-k_2, q_2-k_3\big]  .
\nonumber
\end{align}

\end{document}